Title:

**Application of Particle Swarm Optimization method to On-going Monitoring for estimating vehicle-bridge interaction system**

Author:

**Yamamoto Kyosuke**, Ph.D.(Eng.), Asst. Prof., University of Tsukuba

**Murakami Kakeru**, Graduate student, University of Tsukuba*

**Shin Ryota**, Graduate student, University of Tsukuba

**Okada Yukihiko**, Ph.D., Assc. Prof., University of Tsukuba

* graduated on March 31, 2021

# ABSTRACT

This study proposes a method for estimating the mechanical parameters of vehicles and bridges and the road unevenness, using only vehicle vibration and position data. In the proposed method, vehicle input and bridge vibration are estimated using randomly assumed vehicle and bridge parameters. Then, the road profiles at the front and rear wheels can be determined from the vehicle input and bridge vibration. The difference between the two road profiles is used as the objective function because they are expected to coincide when synchronized. Using the particle swarm optimization (PSO) method, the vehicle's and bridge's parameters and the road unevenness can be estimated by updating the parameters to minimize the objective function. Numerical experiments also verify the applicability of this method. In the numerical experiments, it is confirmed that the proposed method can estimate the vehicle weight with reasonable accuracy, but the accuracy of other parameters is not sufficient. It is necessary to improve the accuracy of the proposed method in the future.

# 1. Introduction

## 1.1 Background

Civil engineers and researchers share a critical awareness of aging bridges' problems and develop vibration-based bridge inspection techniques[1]-[5]. On the other hand, in recent years, maintenance risks are increasing not only for bridges but also for the entire road network. For example, vehicle breakdown accidents and the neglect of deteriorated road pavement are also severe problems. Therefore, a new maintenance management system that constantly diagnoses the entire road network, including vehicles, bridges, and pavements, based on vibration information is considered necessary. However, for safety reasons, rigorous accuracy is required to inspect vehicles and bridges. Thus, this study focuses on developing a primary screening technology that can quickly and cost-effectively diagnose vehicles, bridges, and road surfaces and prompt detailed inspections for those with a high failure probability.

A possible candidate for inspecting the entire road network is to use a traveling vehicle's vibration and position data. This method is called "On-going Monitoring" [6]-[8] in this study. When using this method, the inspector only needs to install the sensors on the monitoring vehicle and does not need to perform complicated multi-point measurements of the bridge. It is also desirable that there is no need to measure the road unevenness using a profiler. **Fig. 1** shows a conceptual diagram of the advantages of on-going monitoring. As shown in the figure, bridge multi-point measurement is labor-intensive because inspectors need to install many sensors on a massive bridge structure. If vehicle monitoring can replace bridge monitoring, the labor-saving effect is very high.

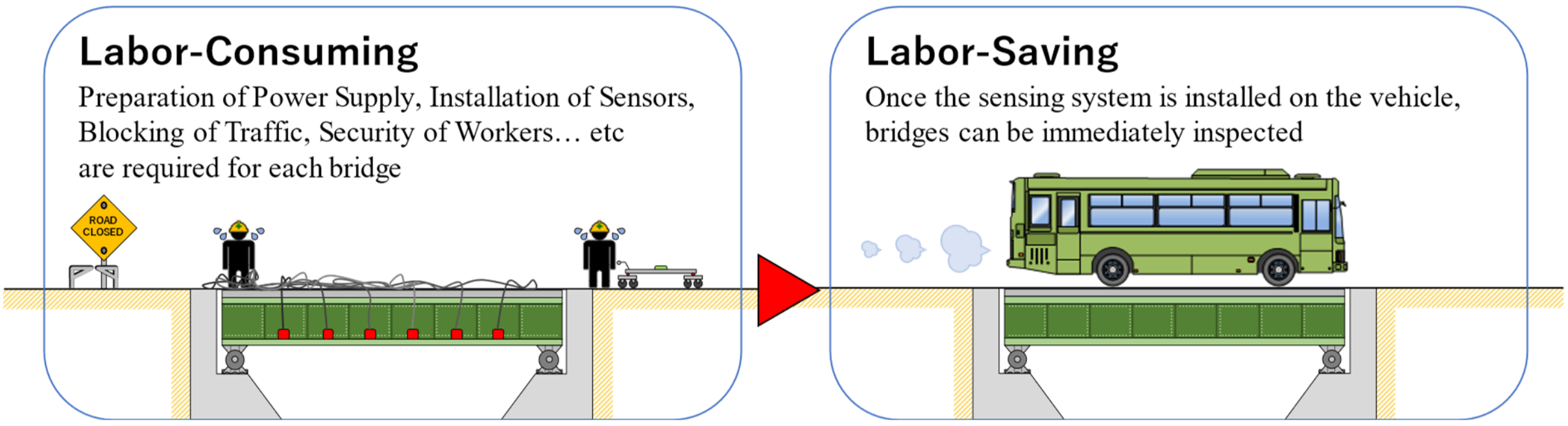

**Fig.1** The concept of "On-going Monitoring"

On-going monitoring has been developed to evaluate road pavements or bridges. The target is either, but not both. For example, Nagayama et al. [9] develop a method to estimate IRI (International Road Index) by vehicle vibration, and Yang et al. [10] propose an "Indirect Approach" to estimate a bridge's natural frequency from vehicle vibration. Thus, it is necessary to integrate both techniques to establish the entire road network screening.

Successful examples of social implementation of vehicle monitoring techniques are mainly found in the former, i.e., road pavement assessment. Since the adoption of IRI as an investment indicator by the World Bank[11], the demand for IRI measurement has increased enormously, especially in developing countries that need financial supports. The monitoring system proposed by Nagayama et al.[9] can collect IRI on a large number of routes quickly, easily, and at a low cost; according to a previous paper published in 2018[12], inspectors can estimate IRI just by setting a smartphone on the vehicle. His system accesses the smartphone's accelerometer and gyrosensor to obtain the vehicle body's acceleration and angular velocity. By substituting these obtained signals into the equation of motion of the vehicle, the road profile can be estimated. However, to solve this equation, the vehicle's mechanical parameters, such as mass, damping, and stiffness, need to be calibrated in advance.

Yang et al.[10] have proposed an "indirect approach" as a bridge monitoring method using vehicle vibration. They experimentally confirm that the bridge's natural frequency can be extracted from the Fourier power spectrum of vehicle vibration[13]. The vehicle vibration includes the bridge response components caused by the vehicle-bridge interaction. According to this paper, if the vehicle's and bridge's natural frequencies are different, the two can be easily distinguished. However, this research concentrates on estimating the bridge vibration frequency and does not yet intend to detect bridge damage or evaluate bridge health. Its estimation accuracy is insufficient to detect damage in bridge structures. K.C. Chang [14] attempted to accurately estimate the bridge's frequency by EMD (Empirical Mode Decomposition). This method can also estimate the mode shapes that are more sensitive to the bridge's local damage than the natural frequencies. Since bridge damage is usually localized, global vibration indices such as natural frequency are inefficient for damage detection. Therefore, research in bridge monitoring often focuses on finding indices highly sensitive to local damage. For example, Nguen et al.[15] numerically show that cracks in beams can be detected from the wavelet coefficients of vehicle vibration. Besides, Xiang et al.[16] applied STFT (Short-Time Fourie's Transform) to the vibration data measured on a vehicle with a shaker to detect the bridge's local damage. This method is called the woodpecker method or the tap-scan method, and although it is not very robust, it shows very high sensitivity under ideal conditions. In other words, these studies show that time-frequency domain indices such as wavelet coefficients and STFT power are hypersensitive to local damage. As the vehicle moves, the time-domain information is transformed into spatial-domain information in on-going monitoring. In other words, since the vehicle position becomes a time function, the bridge's spatial characteristics can be evaluated by analyzing the time domain information. Recently, Malekjafarian[17] proposed the STFDD (Short-Time Frequency Domain Decomposition) method to estimate bridges' mode shapes. The mode shape is a well-known spatial vibration index, and it has high sensitivity. However, the damage detection accuracy of the STFDD method depends on the accuracy of mode shape estimation and is still insufficient.

Based on these studies, the authors designed and proposed SSMA (Spatial Singular Mode Angle) as a damage detection index[18]-[20]. SSMA is a bridge mode shape estimated from vehicle vibration data and shows a high sensitivity to local damage. The studies numerically and experimentally confirm that SSMA is effective for damage detection and can detect even minor bridge damage. However, SSMA only indicates the "existence of damage," not its size, location, magnitude, or type.

When applying the methods mentioned above to the simultaneous vehicle-bridge-road inspection, the lack of system identification is a common issue. The existing on-going monitoring methods for bridge condition assessment calculate only the evaluation index, not the system parameters. The system parameters such as vehicle mass, suspension damping and stiffness, tire weight and stiffness, bridge bending stiffness distribution, mass distribution, and damping are not directly identified. Thus, this study proposes a method to identify vehicles' and bridges' system parameters and estimate the road surface roughness, and it also validates the method by numerical experiments.

## 1.2. The outline of the proposed method

The authors considered that a possible method for identifying the vehicle's and bridge's system parameters and estimating the road surface roughness could be realized by extending the method[9] of Nagayama et al. Their method can estimate the road surface roughness if the vehicle parameters are known. The estimated value should include the bridge vibration component when the vehicle passes over a bridge. The vehicle's system input is referred to as the input profile, the road surface unevenness-derived component included in the input profile as the road profile, and the bridge vibration-derived component as the bridge profile. The bridge profile can be estimated from vehicle weight and vehicle vibration data. The road profile should be estimated by taking the difference between the input profile and the bridge profile. The vehicle's and bridge's system parameters must be known, but are assumed to be random. The road profiles estimated for the front and rear wheels should match if they are spatially synchronized. This means that their difference can be used as the objective function. If the randomly assumed parameters can be updated using this objective function, the desired method can be constructed.

# 2. Numerical Experiment

First, this section explains the numerical experiment of a vehicle traveling on a bridge. The vehicle vibration and position data are measured.

Sensors are mounted on the running vehicle to obtain the vehicle's vertical acceleration vibration and position data. The vehicle's and the bridge's dynamic behaviors can be simulated by numerically solving each equation of motion. The vehicle vibration becomes the loads acting on the bridge, and the deflection of the bridge becomes the vehicle's input profile. This interaction system is called the Vehicle-Bridge Interaction (VBI) system and is nonlinear. The Newmark-beta and Newton-Raphson methods are applied to solve the equations.

In this section, the results of numerical experiments are also presented. The road profile was randomly generated based on actual characteristics. In the numerical experiments, a 10-ton vehicle passes over a bridge with a span of 30 [m] at a speed of 10 [m/s]. Two bridge models are prepared: an intact bridge and a damaged bridge. The intact bridge is assumed to have a homogeneous beam structure with constant flexural rigidity. In contrast, the damaged bridge is assumed to have a 50% reduction in flexural rigidity at the center of the span. Assuming that sensors are installed on the vehicle body directly above the front and rear wheels, the simulated acceleration responses are used as the measured data.

## 2.1. Vehicle-Bridge Interaction System

### (1) Finite Element Model for the bridge

Let "flexural rigidity" and "mass per unit length" of a bridge be $EI$ and $\rho A$, respectively, the equation of motion of the bridge system can be expressed as

$$\rho A\ddot{y}(t,x)+\frac{\partial^2}{\partial x^2}\left(EI\frac{\partial^2 y}{\partial x^2}\right)=\sum_{i=1}^{n}\delta(x-x_i)P_i \tag{1}$$

where $y(t,x)$ denotes the deflection, $t$ represents the time, $x$ represents the position, $x=0$ indicates the bridge's entrance, and $x=L$ indicates the exit. The bridge span length is $L$. The symbol of $(\ddot{\ })$ represents the second-order time derivative. The index $i$ represents each vehicle axle, and $n$ is the number of vehicle axles. The load $P_i$ is the contact force due to the $i$-th axle. The function $\delta(x)$ is Dirac's Delta function, which satisfies

$$\delta(x)=\begin{cases}0 & (\text{at } x\neq 0)\\ \infty & (\text{at } x=0)\end{cases} \tag{2}$$

and

$$\int_{-\infty}^{\infty}\delta(x)f(x)\,\mathrm{d}x=f(0) \tag{3}$$

where $f(x)$ is an arbitrary function. The term of $\delta(x-x_i)P_i$ in Eq. (1) expresses the concentrated load acting on $x=x_i$. Eq. (1) assumes that this bridge structure satisfies the Euler-Bernoulli theorem. **Fig. 2** shows a diagram of the bridge model of Eq. (1).

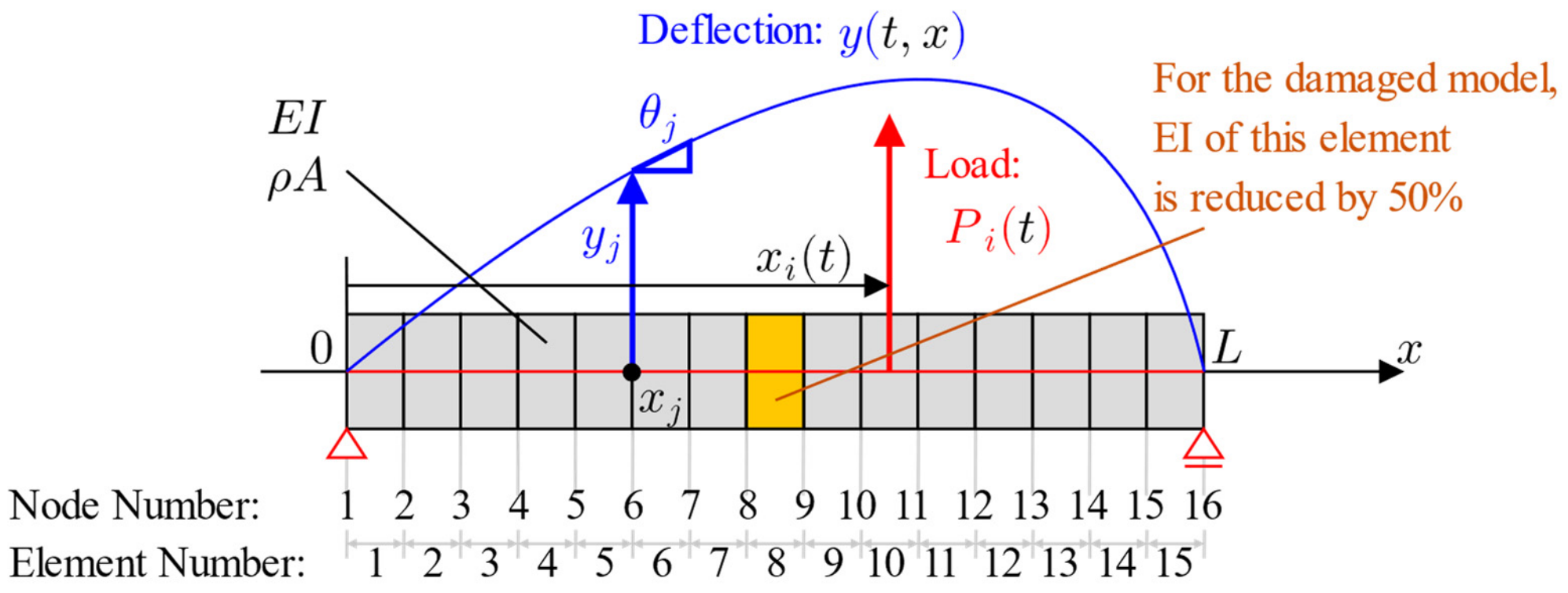

**Fig. 2** The bridge model

FEM (Finite Element Method) is applied to solve Eq. (1) numerically. The finite element formulation is derived by WRM (Weighted Residual Method). As the basis function, Hermite Interpolation Functions are adopted:

$$\begin{aligned}\phi_1(X)&=\frac{1}{4}(X-1)^2(X+2)\\ \phi_2(X)&=\frac{1}{4}(X-1)^2(X+1)\\ \phi_3(X)&=-\frac{1}{4}(X+1)^2(X-2)\\ \phi_4(X)&=\frac{1}{4}(X+1)^2(X-1)\end{aligned} \tag{4}$$

where $X$ represents the normalized local coordinate in each element. Assuming that $j$-th and $(j+1)$-th nodes compose $j$-th beam element, $X=-1$ indicates the position of $j$-th node, and $X=1$ indicates the position of $(j+1)$-th node. The solution $y(t,x)$ can be approximately expressed as

$$\begin{aligned} y(t,x) &= \boldsymbol{y}(t)\cdot\boldsymbol{N}(x) \\ &= \begin{Bmatrix} y_1(t) \\ \theta_1(t) \\ \vdots \\ y_n(t) \\ \theta_n(t) \end{Bmatrix} \cdot \begin{Bmatrix} N_1(x) \\ N_2(x) \\ \vdots \\ N_{2n-1}(x) \\ N_{2n}(x) \end{Bmatrix} \end{aligned} \tag{5}$$

where $\boldsymbol{y}(t)$ is the numerical solution vector, and $\boldsymbol{N}(x)$ is the basis function vector. The component $y_j(t)$ denotes $y(t,x_j)$. which is the bridge deflection at $j$-th node $x_j$. Similarly, when $\theta=\partial y/\partial x$, $\theta_j(t)$ denotes $\theta(t,x_j)$, which is the bridge deflection angle at $j$-th node $x_j$. When $x_j \le x \le x_{j+1}$, the basis function vector becomes

$$N_{2j-2+k}(x) = \begin{cases} \phi_k(X) & (k=1,3) \\ \dfrac{\varDelta x}{2}\phi_k(X) & (k=2,4) \\ 0 & (k=\text{others}) \end{cases} \tag{6}$$

where $\varDelta x$ is the length of $j$-th element, and the global coordinate $x$ can be transformed into the local coordinate $X$ by the following:

$$x = \begin{Bmatrix} x_j \\ \varDelta x/2 \\ x_{j+1} \\ \varDelta x/2 \end{Bmatrix} \cdot \begin{Bmatrix} \phi_1(X) \\ \phi_2(X) \\ \phi_3(X) \\ \phi_4(X) \end{Bmatrix} \tag{7}$$

where $\varDelta x = x_{j+1}-x_j$. Substituting the approximate solution described by Eq. (5) into Eq. (1), the residual is

$$R(x,t) = \rho A\ddot{y}(x,t) + \frac{\partial^2}{\partial x^2}\left(EI\frac{\partial^2 y}{\partial x^2}\right) - \sum_{i=1}^{n}\delta(x-x_i)P_i. \tag{8}$$

If the average of weighted residual always equals zero for any weight function,

$$\begin{aligned} I &= \int_0^L w(x)R(x,t)\mathrm{d}x \\ &= \int_0^L w\rho A\ddot{y}\mathrm{d}x + \int_0^L w\frac{\partial^2}{\partial x^2}\left(EI\frac{\partial^2 y}{\partial x^2}\right)\mathrm{d}x - \sum_{i=1}^{n}\int_0^L w(x)\delta(x-x_i)P_i\mathrm{d}x \\ &= \int_0^L w\rho A\ddot{y}\mathrm{d}x + [w(EIy'')']_0^L - [w'(Ey'')]_0^L + \int_0^L EI\frac{\partial^2 w}{\partial x^2}\frac{\partial^2 y}{\partial x^2}\mathrm{d}x - \sum_{i=1}^{n}\int_0^L w\delta(x-x_i)P_i\mathrm{d}x \\ &= \int_0^L w\rho A\ddot{y}\mathrm{d}x + \int_0^L EI\frac{\partial^2 w}{\partial x^2}\frac{\partial^2 y}{\partial x^2}\mathrm{d}x - \sum_{i=1}^{n} w(x-x_i)P_i = 0. \end{aligned} \tag{9}$$

Substituting the approximate solution $y=\boldsymbol{y}\cdot\boldsymbol{N}$ and $w=\boldsymbol{w}\cdot\boldsymbol{N}$ into Eq. (9),

$$I = \boldsymbol{w}^{\mathrm{T}}\left(\left[\int_0^L \rho A\boldsymbol{N}\boldsymbol{N}^{\mathrm{T}}\mathrm{d}x\right]\ddot{\boldsymbol{y}} + \left[\int_0^L EI\frac{\partial^2\boldsymbol{N}}{\partial x^2}\frac{\partial^2\boldsymbol{N}^{\mathrm{T}}}{\partial x^2}\mathrm{d}x\right]\boldsymbol{y} - \sum_{i=1}^{n}\boldsymbol{N}(x-x_i)P_i\right) = 0. \tag{10}$$

The condition that the average of weighted residual is zero for any weight $\boldsymbol{w}$ can be expressed as

$$\left[\int_0^L \rho A \boldsymbol{N}\boldsymbol{N}^{\mathrm{T}} \mathrm{d}x\right] \ddot{\boldsymbol{y}} + \left[\int_0^L EI \frac{\partial^2 \boldsymbol{N}}{\partial x^2} \frac{\partial^2 \boldsymbol{N}^{\mathrm{T}}}{\partial x^2} \mathrm{d}x\right] \boldsymbol{y} - \sum_{i=1}^{n} \boldsymbol{N}(x - x_i) P_i = \boldsymbol{0}. \tag{11}$$

When $\mathbf{M}_b = \left[\int_0^L \rho A \boldsymbol{N}\boldsymbol{N}^{\mathrm{T}} \mathrm{d}x\right]$, $\mathbf{K}_b = \left[\int_0^L EI \frac{\partial^2 \boldsymbol{N}}{\partial x^2} \frac{\partial^2 \boldsymbol{N}^{\mathrm{T}}}{\partial x^2} \mathrm{d}x\right]$, $\boldsymbol{L} = [\boldsymbol{N}(x - x_1) \ldots\ \boldsymbol{N}(x - x_n)]$ and $\boldsymbol{P} = \begin{Bmatrix} P_1 \\ \vdots \\ P_n \end{Bmatrix}$, Eq.(10) can be rewritten in the form of finite element formula:

$$\mathbf{M}_b \ddot{\boldsymbol{y}} + \mathbf{K}_b \boldsymbol{y} = \mathbf{L}\boldsymbol{P}. \tag{12}$$

For numerical stability, the Raleigh damping is also introduced.

$$\mathbf{M}_b \ddot{\boldsymbol{y}} + \mathbf{C}_b \dot{\boldsymbol{y}} + \mathbf{K}_b \boldsymbol{y} = \mathbf{L}\boldsymbol{P} \tag{13}$$

where $\mathbf{C}_b = \alpha_c \mathbf{M}_b + \beta_c \mathbf{K}_b$.

## (2) Rigid-Body-Spring Model for Vehicle

**Fig. 3** shows the RBSM (Rigid Body Spring Model) of the vehicle system. In this model, $m_s$ is the mass of the vehicle body and $c_{si}$, $k_{si}$, $d_i$, $m_{ui}$ and $k_{ui}$ represent the suspension damping, the suspension stiffness, the distance from the gravity point, the unsprung mass, and the tire stiffness of $i$-th axle, respectively. The vehicle's dynamic behavior can be obtained as the simultaneous solution of four equations: the translational and rotational equations of motion given for the vehicle body, and the equations of motion of two unsprung-mass. The translational equation of motion of $m_s$ can be expressed as

$$m_s \ddot{z}_G = -k_{s1}(z_{s1} - z_{u1}) - k_{s2}(z_{s2} - z_{u2}) - c_{s1}(\dot{z}_{s1} - \dot{z}_{u1}) - c_{s2}(\dot{z}_{s2} - \dot{z}_{u2}) \tag{14}$$

where $z_G$ is the vertical displacement of the vehicle body at the gravity point, and $z_{si}$ and $z_{ui}$ represent the vertical displacement of the vehicle body and the unsprung mass at $i$-th axle. The rotational equation of motion of $m_s$ can be given by

$$I_s \ddot{\theta}_G = -k_{s1} d_1 (z_{s1} - z_{u1}) + k_{s2} d_2 (z_{s2} - z_{u2}) - c_{s1} d_1 (\dot{z}_{s1} - \dot{z}_{u1}) + c_{s2} d_2 (\dot{z}_{s2} - \dot{z}_{u2}) \tag{15}$$

where $I_s$ and $\theta_G$ are the inertia moment and the rotation angle, respectively. $z_G$ and $\theta_G$ can be represented by $z_{s1}$ and $z_{s2}$ as

$$z_G = \frac{d_2}{d_1 + d_2} z_{s1} + \frac{d_1}{d_1 + d_2} z_{s2} \tag{16}$$

$$\theta_G = \frac{z_{s1} - z_{s2}}{d_1 + d_2} \tag{17}$$

If the center of the rotation coincides with the gravity point, it is known that

$$I_s = m_s d_1 d_2. \tag{18}$$

The equation of motion of each unsprung mass point is expressed as

$$m_{ui} \ddot{z}_{ui} = k_{si}(z_{si} - z_{ui}) + c_{si}(\dot{z}_{si} - \dot{z}_{ui}) - k_{ui}(z_{ui} - u_i) \tag{19}$$

where $u_i$ is the input profile at $i$-th axle. The input profile is the road profile $r_i(t)$ plus the bridge profile $\tilde{y}_i(t)$. The road profile $r_i(t)$ can be obtained from the road unevenness $R(x)$, which is a spatial function. The bridge profile is the bridge deflection at the position of $i$-th axle ($x = x_i(t)$). Thus,

$$\begin{aligned} u_i(t) &= r_i(t) + \tilde{y}_i(t) \\ &= R\big(x_i(t)\big) + \boldsymbol{y} \cdot \boldsymbol{N}\big(x_i(t)\big). \end{aligned} \tag{20}$$

The input profile vector $\boldsymbol{u}(t) = [u_1(t) \ \cdots u_n(t)]^{\mathrm{T}}$ can be expressed as

$$\boldsymbol{u}(t) = \boldsymbol{r}(t) + \mathbf{L}(t)^{\mathrm{T}}\boldsymbol{y}(t). \tag{21}$$

Substituting Eqs. (16) and (17) into Eqs. (14) and (15), they become

$$\begin{aligned} m_{s1}\ddot{z}_{s1} &= -k_{s1}(z_{s1} - z_{u1}) - c_{s1}(\dot{z}_{s1} - \dot{z}_{u1}) \\ m_{s2}\ddot{z}_{s2} &= -k_{s2}(z_{s2} - z_{u2}) - c_{s2}(\dot{z}_{s2} - \dot{z}_{u2}) \end{aligned} \tag{22}$$

where $m_{s1} = \frac{d_2}{d_1+d_2}m_s$ and $m_{s2} = \frac{d_1}{d_1+d_2}m_s$. This means that the Half-Car model is equivalent to two Quarter-Car models. The system parameter matrices and input/output vectors of the vehicle, which are used for summarizing Eq. (14), (15), and (19), can be given by

$$\mathbf{M}_v = \begin{bmatrix} \dfrac{d_2 m_s}{d_1+d_2} & \dfrac{d_1 m_s}{d_1+d_2} & & \\ \dfrac{I_s}{d_1+d_2} & \dfrac{-I_s}{d_1+d_2} & & \\ & & m_{u1} & \\ & & & m_{u2} \end{bmatrix} \ (23) \qquad \boldsymbol{z} = \begin{Bmatrix} z_{s1} \\ z_{s2} \\ z_{u1} \\ z_{u2} \end{Bmatrix} \ (24) \qquad \boldsymbol{f}_v = \begin{Bmatrix} 0 \\ 0 \\ k_{u1}u_1 \\ k_{u2}u_2 \end{Bmatrix} \ (25)$$

$$\mathbf{C}_v = \begin{bmatrix} c_{s1} & c_{s2} & -c_{s1} & -c_{s2} \\ c_{s1}d_1 & -c_{s2}d_2 & -c_{s1}d_1 & c_{s2}d_2 \\ -c_{s1} & & c_{s1} & \\ & -c_{s2} & & c_{s2} \end{bmatrix} \ (26) \quad \mathbf{K}_v = \begin{bmatrix} k_{s1} & k_{s2} & -k_{s1} & -k_{s2} \\ k_{s1}d_1 & -k_{s2}d_2 & -k_{s1}d_1 & k_{s2}d_2 \\ -k_{s1} & & k_{s1}+k_{u1} & \\ & -k_{s2} & & k_{s2}+k_{u2} \end{bmatrix} \ (27)$$

Assuming

$$\mathbf{F}_v = \begin{bmatrix} 0 & 0 \\ 0 & 0 \\ k_{u1} & 0 \\ 0 & k_{u2} \end{bmatrix} \tag{28}$$

the load $\boldsymbol{f}_v$ can be rewritten in

$$\boldsymbol{f}_v = \mathbf{F}_v \boldsymbol{u} \tag{29}$$

Then, the equation of motion of the vehicle can be summarized into

$$\mathbf{M}_v\ddot{\boldsymbol{z}} + \mathbf{C}_v\dot{\boldsymbol{z}} + \mathbf{K}_v\boldsymbol{z} = \boldsymbol{f}_v. \tag{30}$$

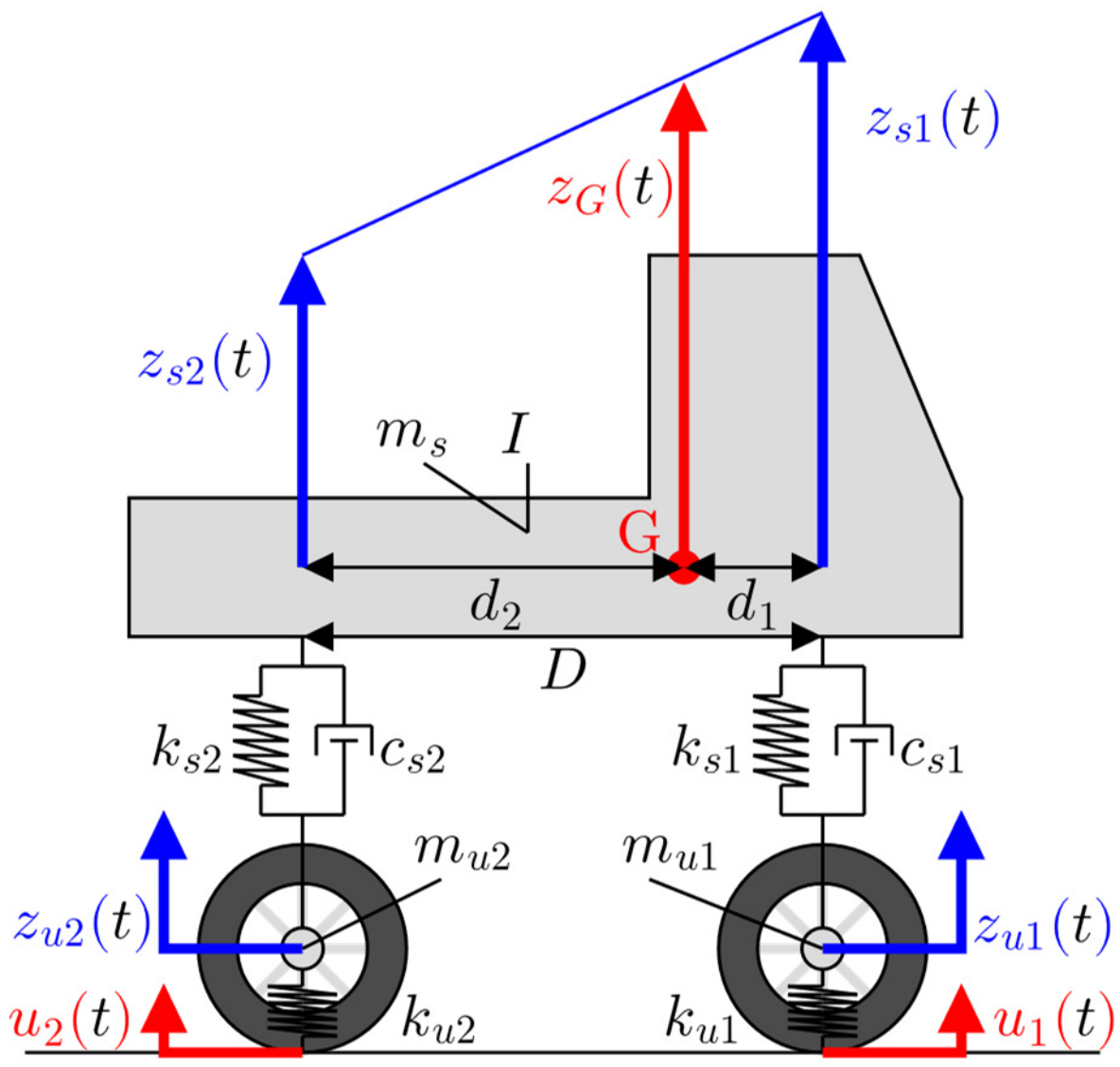

**Fig. 3** Rigid-Body-Spring model for the vehicle

## (3) The contact forces

The contact force $P_i(t)$ is the restoring force of the tire $k_{ui}$.

$$P_i = k_{ui}(z_{ui} + z_{ui}^0 - u_i) \tag{31}$$

where $z_{ui}^0$ is the elongation of the tire spring $k_{ui}$ as it changes from the natural length to the balanced length. In the equation of motion of the unsprung mass, the gravity term is canceled. When the vehicle is in the static state of the equilibrium, the equation of equilibrium of the vehicle can be expressed as

$$\begin{aligned} &-m_s g - k_{s1}(z_{s1}^0 - z_{u1}^0) - k_{s2}(z_{s2}^0 - z_{u2}^0) = 0 \\ &-k_{s1} d_1 (z_{s1}^0 - z_{u1}^0) + k_{s2} d_2 (z_{s2}^0 - z_{u2}^0) = 0 \\ &-m_{ui} g + k_{si}(z_{si}^0 - z_{ui}^0) - k_{ui} z_{ui}^0 = 0 \end{aligned} \tag{32}$$

where $z_{si}^0$ represents the elongation of the suspension spring $k_{si}$ as it changes from the natural length to the balanced length. $g$ is the gravity acceleration. From Eqs. (32), the following equations are available.

$$\begin{aligned} k_{u1} z_{u1}^0 &= -\frac{m_s d_2}{d_1 + d_2} g - m_{u1} g \\ k_{u2} z_{u2}^0 &= -\frac{m_s d_1}{d_1 + d_2} g - m_{u2} g \end{aligned} \tag{33}$$

Thus, substituting Eqs. (19), (22) and (33) into Eq. (31), the contact force $P_i$ becomes

$$\begin{aligned} P_i &= k_{ui}(z_{ui} + z_{ui}^0 - u_i) \\ &= k_{ui}(z_{ui} - u_i) - \left(1 - \frac{d_i}{d_1 + d_2}\right) m_s g - m_{ui} g \\ &= -m_{si}(g + \ddot{z}_{si}) - m_{ui}(g + \ddot{z}_{ui}) \end{aligned} \tag{34}$$

## (4) Application of Newmark-$\beta$ method

Newmark Beta method is adopted to discretize the time derivative terms to simulate the vehicle's and bridge's dynamic behaviors. When the equation of motion is given by

$$\mathbf{M}\ddot{\boldsymbol{\eta}}(t)+\mathbf{C}\dot{\boldsymbol{\eta}}(t)+\mathbf{K}\boldsymbol{\eta}(t)=\boldsymbol{\xi}(t) \tag{35}$$

where $\mathbf{M}$, $\mathbf{C}$ and $\mathbf{K}$ are the mass, damping and stiffness matrices, respectively, and where $\boldsymbol{\eta}(t)$ and $\boldsymbol{\xi}(t)$ are the displacement and load, respectively. The Newmark Beta method assumes

$$\begin{aligned}\dot{\boldsymbol{\eta}}(t+\Delta t)&=\dot{\boldsymbol{\eta}}(t)+\Delta t\big((1-\gamma)\ddot{\boldsymbol{\eta}}(t)+\gamma\ddot{\boldsymbol{\eta}}(t+\Delta t)\big)\\ \boldsymbol{\eta}(t+\Delta t)&=\boldsymbol{\eta}(t)+\Delta t\dot{\boldsymbol{\eta}}(t)+\Delta t^2\left(\left(\frac{1}{2}-\beta\right)\ddot{\boldsymbol{\eta}}(t)+\beta\ddot{\boldsymbol{\eta}}(t+\Delta t)\right)\end{aligned} \tag{36}$$

where $\gamma$ and $\beta$ are the coefficients of this method. Usually, $\gamma=1/2$ and $\beta=1/4$ or $1/6$. Eq. (35) can be rewritten in

$$\mathbf{M}\ddot{\boldsymbol{\eta}}(t+\Delta t)+\mathbf{C}\dot{\boldsymbol{\eta}}(t+\Delta t)+\mathbf{K}\boldsymbol{\eta}(t+\Delta t)=\boldsymbol{\xi}(t+\Delta t) \tag{37}$$

Substituting Eqs. (36) into Eq. (37), the following equation is available.

$$\mathbf{A}\ddot{\boldsymbol{\eta}}(t+\Delta t)=\boldsymbol{b} \tag{38}$$

where the global matrix is

$$\mathbf{A}=[\mathbf{M}+\gamma\mathbf{C}+\beta\mathbf{K}] \tag{39}$$

and the RHS (right-hand side) term is given by

$$\boldsymbol{b}=\left\{\boldsymbol{\xi}(t+\Delta t)-\mathbf{C}\big(\dot{\boldsymbol{\eta}}(t)+\Delta t(1-\gamma)\ddot{\boldsymbol{\eta}}(t)\big)-\mathbf{K}\left(\boldsymbol{\eta}(t)+\Delta t\dot{\boldsymbol{\eta}}(t)+\Delta t^2\left(\frac{1}{2}-\beta\right)\ddot{\boldsymbol{\eta}}(t)\right)\right\} \tag{40}$$

If $\boldsymbol{\eta}(t)$, $\dot{\boldsymbol{\eta}}(t)$, $\ddot{\boldsymbol{\eta}}(t)$ and $\boldsymbol{\xi}(t+\Delta t)$ are known, $\ddot{\boldsymbol{\eta}}(t+\Delta t)$ can be solved by Eq. (38).

For the vehicle system, $\mathbf{M}_v$, $\mathbf{C}_v$, $\mathbf{K}_v$, $\boldsymbol{z}(t)$, and $\boldsymbol{f}_v(t)$ are corresponding to $\mathbf{M}$, $\mathbf{C}$, $\mathbf{K}$, $\boldsymbol{\eta}(t)$ and $\boldsymbol{\xi}(t)$ of Eq. (37), respectively. On the other hand, for the bridge system, $\mathbf{M}_b$, $\mathbf{C}_b$, $\mathbf{K}_b$, $\boldsymbol{y}(t)$, and $\mathbf{L}(t)\boldsymbol{P}(t)$ are corresponding to the same ones.

### (5) Iterative Calculation by Newton-Raphson method

**Fig. 4** shows the interaction between the vehicle and bridge models. To simulate the vehicle vibration $\boldsymbol{z}(t)$, the bridge vibration $\boldsymbol{y}(t)$ must be known. On the other hand, to determine the bridge vibration $\boldsymbol{y}(t)$, the vehicle vibration $\boldsymbol{z}(t)$ must be given. From this point of view, VBI system is a nonlinear system.

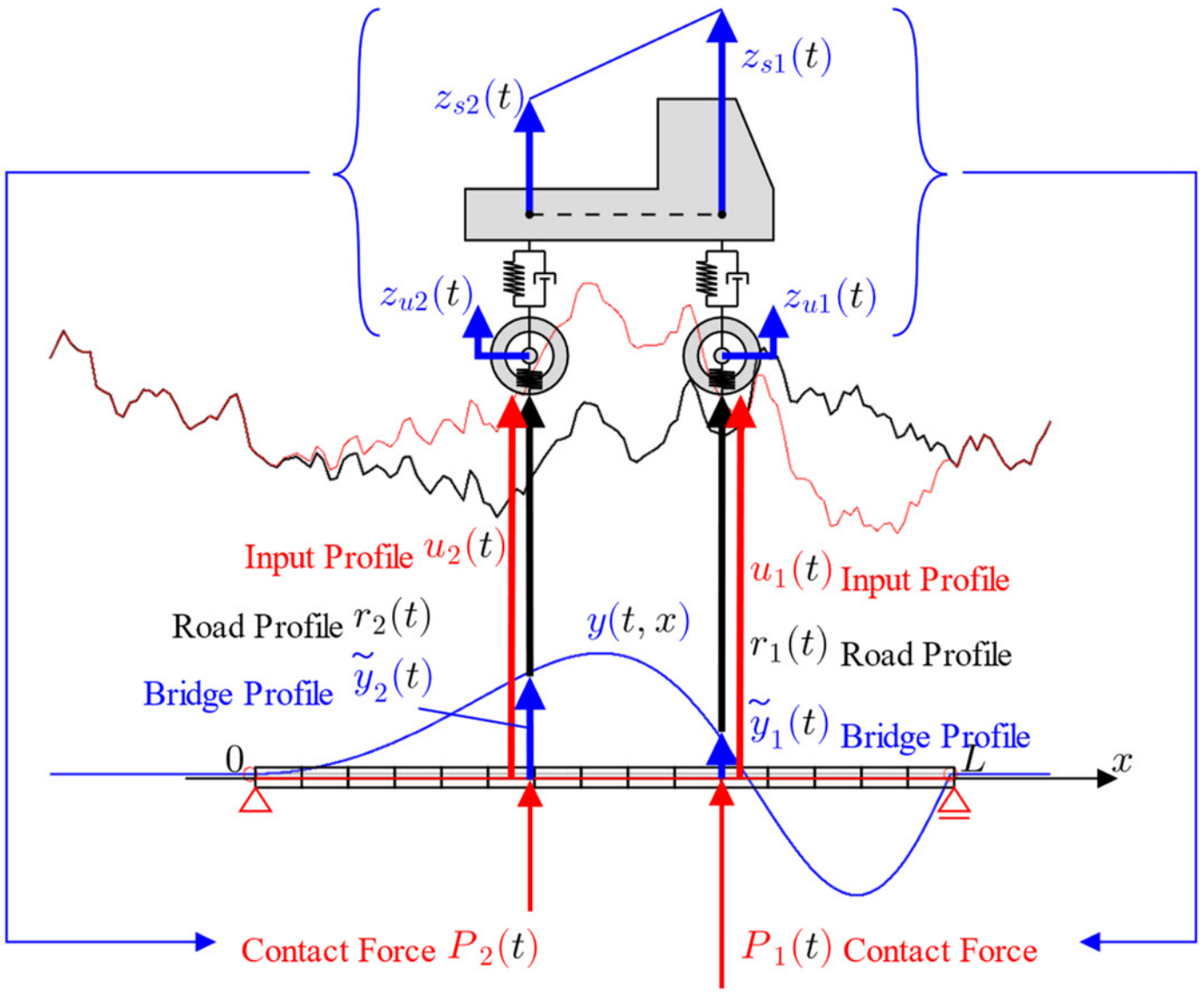

**Fig. 4** VBI (Vehicle-Bridge Interaction) system

When applying Newmark-$\beta$ method to the equation of motion of the vehicle, it becomes

$$\mathbf{A}\ddot{\boldsymbol{z}}(t+\Delta t)=\boldsymbol{b}. \tag{41}$$

where $\mathbf{A}=[\mathbf{M}_v+\gamma\mathbf{C}_v+\beta\mathbf{K}_v]$. The RHS term can be expressed as

$$\boldsymbol{b}=\left\{\mathbf{F}_b\boldsymbol{u}(t+\Delta t)-\mathbf{K}_v\boldsymbol{z}(t)-[\mathbf{C}_v+\Delta t\mathbf{K}_v]\dot{\boldsymbol{z}}(t)-\left[\Delta t(1-\gamma)\mathbf{C}_v+\Delta t^2\left(\frac{1}{2}-\beta\right)\mathbf{K}_v\right]\ddot{\boldsymbol{z}}(t)\right\} \tag{42}$$

The input profile $\boldsymbol{u}$ includes the bridge vibration component $\mathbf{L}^{\mathrm{T}}\boldsymbol{y}(t)$. Because the bridge vibration is excited by the vehicle vibration, the RHS term $\boldsymbol{b}$ is a function of $\ddot{\boldsymbol{z}}(t+\Delta t)$. This means that the RHS term of Eq. (41) is a nonlinear term of $\ddot{\boldsymbol{z}}(t+\Delta t)$.

$$\mathbf{A}\ddot{\boldsymbol{z}}(t+\Delta t)=\boldsymbol{b}\big(\ddot{\boldsymbol{z}}(t+\Delta t)\big). \tag{43}$$

To solve this, the Newton-Raphson method can be applied. First, by assuming that $\ddot{\boldsymbol{z}}=\mathbf{0}$ to calculate the RHS term, which means that input profile is only road profile ($\boldsymbol{u}=\boldsymbol{r}$), a tentative solution $\ddot{\boldsymbol{z}}^{(1)}=\mathbf{A}^{-1}\boldsymbol{b}(\mathbf{0})$ is available. Substituting $\ddot{\boldsymbol{z}}^{(1)}$ instead of $\ddot{\boldsymbol{z}}$ into the equation of motion of the bridge (Eq.(13)), a tentative bridge vibration $\boldsymbol{y}^{(1)}$can also be calculated. Thus, the input profile can be updated ($\boldsymbol{u}=\boldsymbol{r}+\mathbf{L}^{\mathrm{T}}\boldsymbol{y}^{(1)}$), and the different vehicle's vibration $\ddot{\boldsymbol{z}}^{(2)}$ can be obtained from the updated RHS term ($\boldsymbol{b}(\boldsymbol{z}^{(1)})$). When repeating the above operation, the following equation can be available.

$$\ddot{\boldsymbol{z}}^{(k+1)}=\mathbf{A}^{-1}\boldsymbol{b}\big(\ddot{\boldsymbol{z}}^{(k)}\big). \tag{44}$$

If $k$ is large enough, $\ddot{\boldsymbol{z}}^{(k+1)}$ converges to the vehicle vibration $\ddot{\boldsymbol{z}}$. This process is a convergence calculation method that finds a vehicle vibration from ignoring the bridge component first, obtains a bridge vibration from the vehicle vibration, updates the input profiles, and repeats until the vehicle vibration converges.

### (6) The measured data

This study assumes two sensors (accelerometers) on the vehicle body at the positions of the front and rear axles. Thus, the measured data are the vertical acceleration vibrations $\ddot{z}_{s1}(t)$ and $\ddot{z}_{s2}(t)$.

## 2.2. Input Parameters of Numerical Simulation

The vehicle and bridge parameters for numerical simulation are shown in **Table 1** and **Table 2**, respectively. The vehicle is 10-ton 2-axle car, of which total mass is 9530[kg] and runs at the speed of 10[m/s] (=36[km/h]). The natural frequencies of this vehicle are 1.123[Hz], 2.035[Hz], 12.63[Hz] and 16.83[Hz]. The vehicle's suspension and tire characteristics are determined to be the same as those used by Nagayama et al.[12], while the vehicle weight is increased in order to shake the bridge sufficiently. According to the existing studies[8], the bridge parameters are also determined. There are two bridge models: intact and damaged. The damaged bridge has 50% less flexural rigidity $EI$ at the mid-span element. The natural frequencies of the bridge for the intact case are 3.305 [Hz], 13.43 [Hz], 31.05 [Hz], 57.24 [Hz]…etc., while those of the damaged bridge are 3.105 [Hz], 13.42 [Hz], 29.40 [Hz], 56.99 [Hz]…etc. The road profile is shown in **Fig. 6**. Because the bridge length is 30[m], the vehicle passes through the span in 3.0[s]. This simulation starts the calculation when the vehicle is 10[m] before the bridge, and it ends when the vehicle is 20[m] past.

**Table 1** The vehicle model parameters

(a) Sprung-mass, suspension characteristics and sensor position

| $m_s$ [kg] | $c_{s1}$ [kg/s] | $c_{s2}$ [kg/s] | $k_{s1}$ [N/m] | $k_{s2}$ [N/m] |
|---|---|---|---|---|
| 8310 | 24200 | 29000 | 456000 | 410000 |

(b) Axle position, unsprung-mass and tire characteristics

| $d_1$ [m] | $d_2$ [m] | $m_{u1}$ [kg] | $m_{u2}$ [kg] | $k_{u1}$ [N/m] | $k_{u2}$ [N/m] |
|---|---|---|---|---|---|
| 1.215 | 3.185 | 469.0 | 751.0 | 4790000 | 4310000 |

**Table 2** The bridge model parameters

| $\rho A$ [kg/m] | $EI$ [Nm$^3$] | $\alpha_c$ [kg/s] | $\beta_c$ [N/m] | $L$ [m] | $\Delta L$ [m] |
|---|---|---|---|---|---|
| 4400 | $1.560\times10^{10}$ | 0.7024 | 0.005200 | 30.00 | 2.000 |

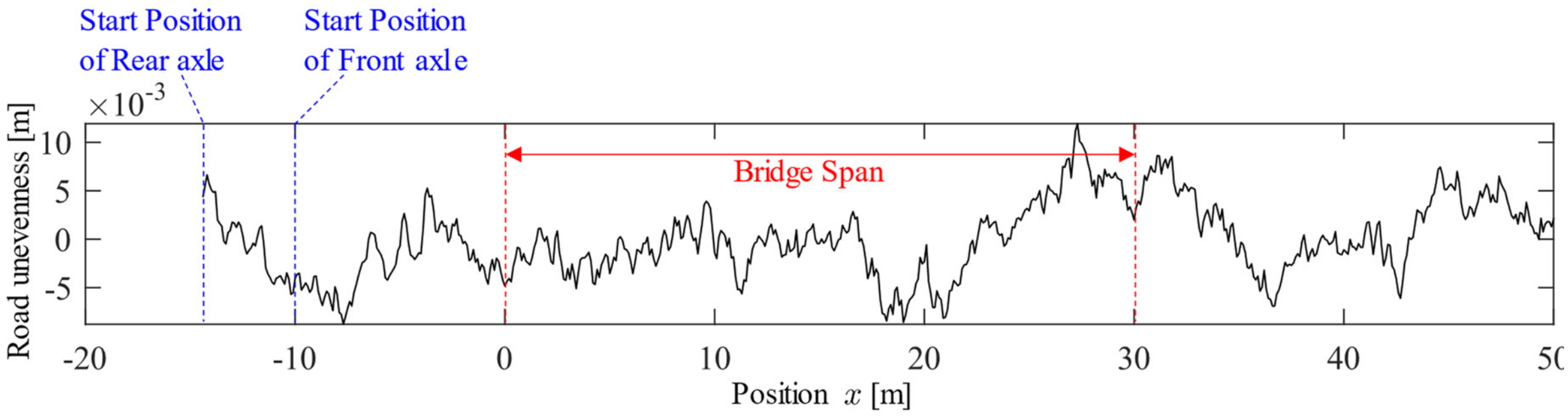

**Fig. 6** The road unevenness

As the simulation parameters, the coefficients of the Newmark-Beta method $\gamma$ and $\beta$, the time increment $\Delta t$, the simulation time $T$, and the maximum error of convergence calculation $\varepsilon_{\max}$ are shown in **Table 3**. The coefficients of the Newmark-Beta method $\gamma$ and $\beta$ are shown in Eq. (36). When $\beta = 1/6$, numerical integration is more accurate but unstable. The parameter setting of $\beta = 1/4$ makes the numerical simulation stable. The convergence of iterative calculation in Eq. (44) is judged by the following equation:

$$\varepsilon = \frac{|\boldsymbol{Z}^{k+1} - \boldsymbol{Z}^{k}|}{|\boldsymbol{Z}^{k+1}|} \leq \varepsilon_{\max} \tag{45}$$

where $\boldsymbol{Z}^k = [\boldsymbol{z}^k(\Delta t)\ \boldsymbol{z}^k(2\Delta t)\ \boldsymbol{z}^k(3\Delta t)\ \cdots]$ is the data matrix of the vehicle vibrations. The operator $|\quad|$ indicates the 2-norm of a matrix.

**Table 3** The parameters of numerical simulation

| $\gamma$ | $\beta$ | $\Delta t$ [s] | $T$ [s] | $\varepsilon_{\max}$ |
|---|---|---|---|---|
| $\frac{1}{2}$ | $\frac{1}{4}$ | 0.001 | 6 | $10^{-6}$ |

## 2.3. Simulated Responses of VBI system

The simulated vibration data of the vehicle are shown in **Fig. 7**. Noises with 0%, 10%, and 35% RMS (root mean square) of the data are added to consider uncertain factors.

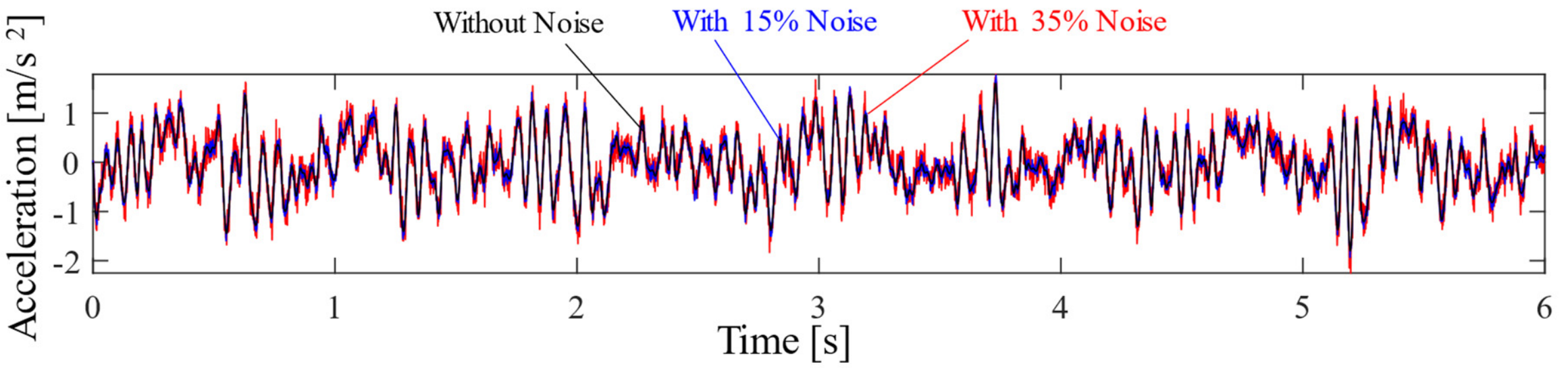

(a) The vertical vibration at the front axle: $\ddot{z}_{s1}(t)$

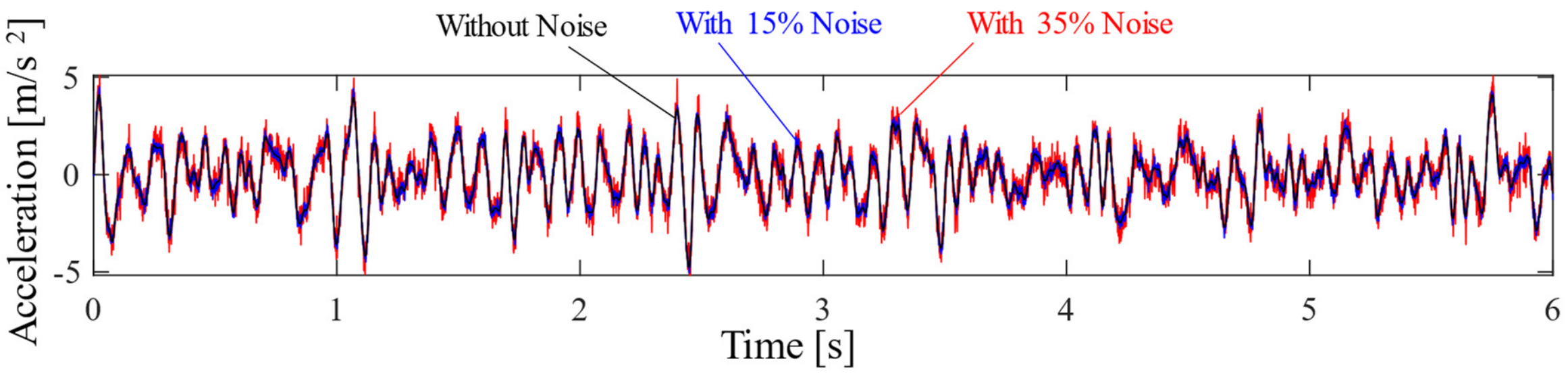

(b) The vertical vibration at the rear axle: $\ddot{z}_{s2}(t)$

**Fig. 7** The simulated vibrations of the vehicle

**Fig. 8** shows the observation point on the bridge, and **Fig. 9** shows the bridge vibration at the mid-span.

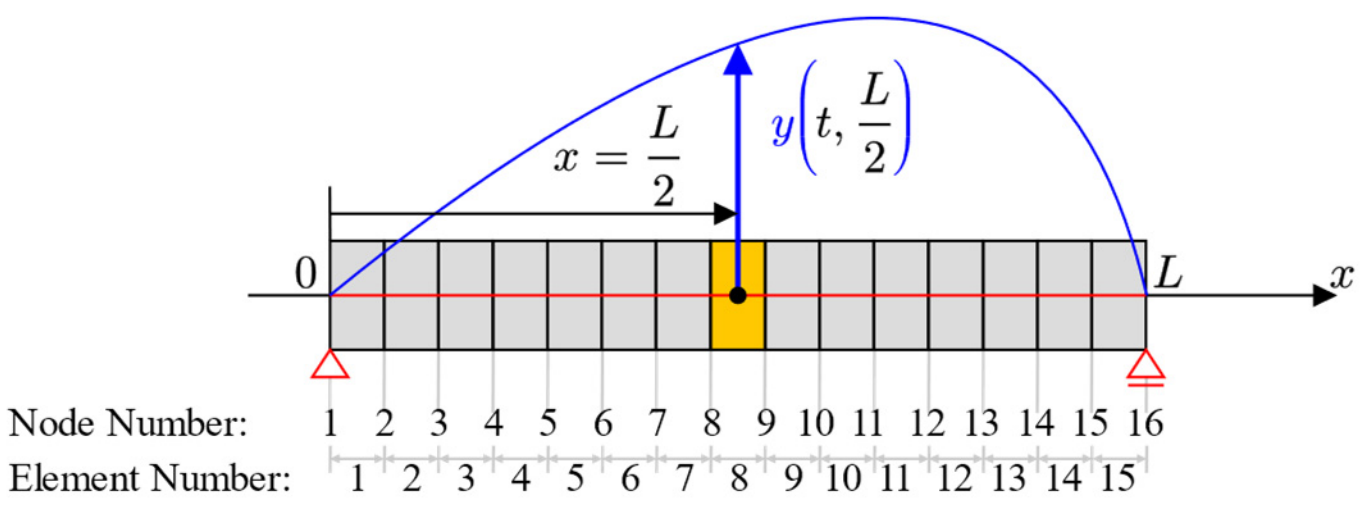

**Fig. 8** The observation point on the bridge

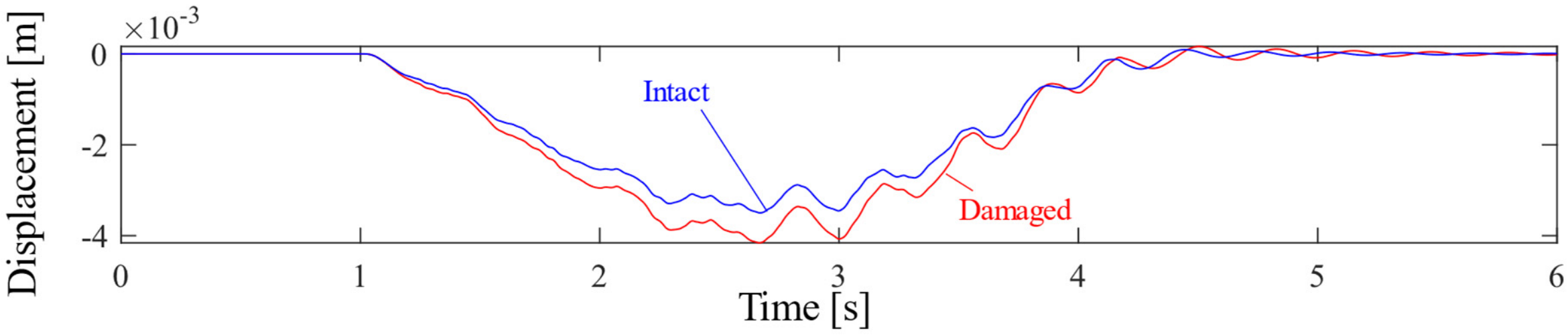

(a) Displacement $y(t, \frac{L}{2})$

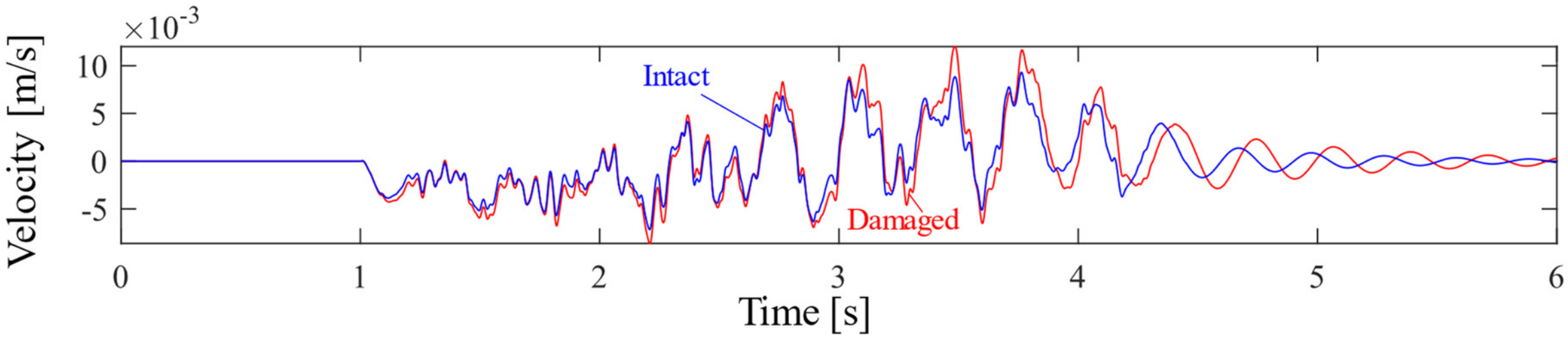

(b) Velocity vibration $\dot{y}(t, \frac{L}{2})$

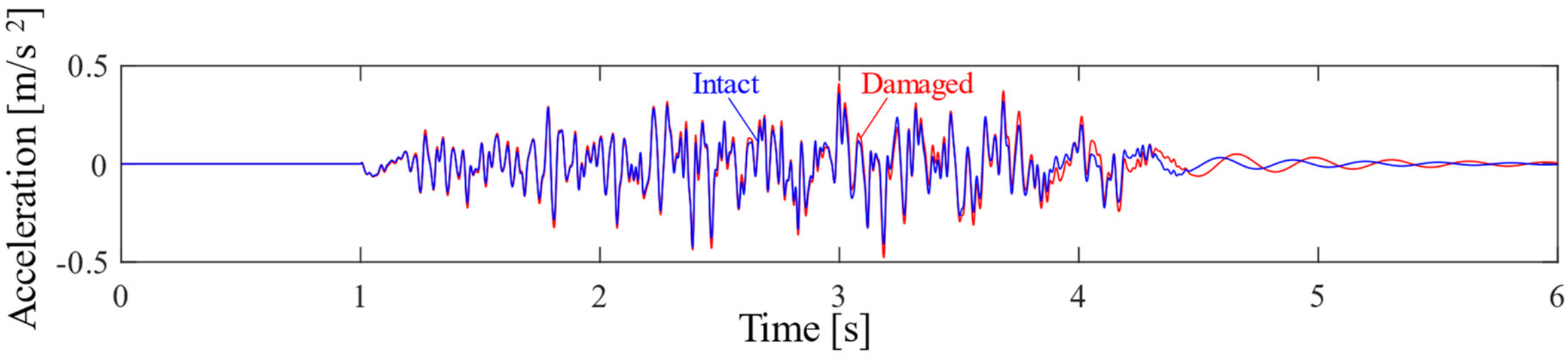

(c) Acceleration vibration $\ddot{y}(t, \frac{L}{2})$

**Fig. 9** The simulated bridge deflection at the mid-span

# 3. System Identification method

The numerical experiment simulates the measured data $\ddot{z}_{s1}(t)$ and $\ddot{z}_{s2}(t)$. The proposed method identifies all system parameters and also estimates the road profile $\boldsymbol{r}(t)$, under the condition that the total weight of the vehicle $M = m_s + m_{u1} + m_{u2}$ and the distance between the front and rear axles $D = d_1 + d_2$ are known. The identified system parameters are the vehicle parameters ($m_{si}$, $m_{ui}$, $k_{si}$, $c_{si}$, $k_{ui}$) and the bridge parameters ($\rho A$, $EI$, $\alpha_c$ and $\beta_c$). This section explains the process of this method.

## 3.1. State Space Model of the vehicle

The vehicle vibrations $\boldsymbol{z}(t)$, $\dot{\boldsymbol{z}}(t)$ and $\ddot{\boldsymbol{z}}(t)$, which are derived from the equation of motion, are summarized in a state variable $\bar{\boldsymbol{Z}}(t)$. On the other hand, the measured data $\ddot{z}_{s1}$ and $\ddot{z}_{s2}$ consist of an observation variable $\bar{\boldsymbol{s}}(t)$. It is noted that the observation variable $\bar{\boldsymbol{s}}(t)$ should be distinguished from the state variable $\bar{\boldsymbol{Z}}(t)$. In this study, according to Nagayama et al.[**], the extended state variable $\boldsymbol{Z}(t)$, which is composed of the input profiles $\boldsymbol{u}(t)$ and $\dot{\boldsymbol{u}}(t)$ as well as the vehicle vibrations, is adopted. The observation variable $\bar{\boldsymbol{s}}(t)$ is also extended by adding $z_{s1}(t)$ and $z_{s2}(t)$ as components. The extended observation variable is represented by $\boldsymbol{s}(t)$.

The state-space representation of Eq. (30), the equation of motion of the vehicle, can be given by the following equations.

$$\dot{\boldsymbol{Z}}(t) = \mathbf{V}\boldsymbol{Z}(t) + \boldsymbol{\omega}(t) \tag{46}$$

$$\boldsymbol{s}(t) = \mathbf{H}\boldsymbol{Z}(t) + \boldsymbol{\epsilon}(t) \tag{47}$$

where

$$\mathbf{V} = \begin{bmatrix} \mathbf{O}^{4\times4} & \mathbf{I}^{4\times4} & \mathbf{O}^{4\times2} & \mathbf{O}^{2\times2} \\ -\mathbf{M}_v^{-1}\mathbf{K}_v & -\mathbf{M}_v^{-1}\mathbf{C}_v & \mathbf{M}_v^{-1}\mathbf{F}_v & \mathbf{O}^{2\times2} \\ \mathbf{O}^{2\times4} & \mathbf{O}^{2\times4} & \mathbf{I}^{2\times2} & \mathbf{O}^{2\times2} \\ \mathbf{O}^{2\times4} & \mathbf{O}^{2\times4} & \mathbf{O}^{2\times2} & \mathbf{O}^{2\times2} \end{bmatrix} \quad (48) \qquad \boldsymbol{Z}(t) = \begin{Bmatrix} \boldsymbol{z}(t) \\ \dot{\boldsymbol{z}}(t) \\ \boldsymbol{u}(t) \\ \dot{\boldsymbol{u}}(t) \end{Bmatrix} \quad (49) \qquad \boldsymbol{s}(t) = \begin{Bmatrix} \ddot{z}_{s1}(t) \\ \ddot{z}_{s2}(t) \\ z_{s1}(t) \\ z_{s2}(t) \end{Bmatrix} \quad (50)$$

$$\mathbf{H} = \begin{bmatrix} -\frac{k_{s1}}{m_s} & 0 & \frac{k_{s1}}{m_s} & 0 & -\frac{c_{s1}}{m_s} & 0 & \frac{c_{s1}}{m_s} & 0 & 0 & 0 & 0 & 0 \\ 0 & -\frac{k_{s2}}{m_s} & 0 & \frac{k_{s2}}{m_s} & 0 & -\frac{c_{s2}}{m_s} & 0 & \frac{c_{s2}}{m_s} & 0 & 0 & 0 & 0 \\ 1 & 0 & 0 & 0 & 0 & 0 & 0 & 0 & 0 & 0 & 0 & 0 \\ 0 & 1 & 0 & 0 & 0 & 0 & 0 & 0 & 0 & 0 & 0 & 0 \end{bmatrix} \tag{51}$$

where $\mathbf{O}$ and $\mathbf{I}$ denote the zero and unit matrices, each upper-right subscript represents matrix size, system noise is $\boldsymbol{\omega}(t)$, and observation noise is $\boldsymbol{\epsilon}(t)$. The displacements components $z_{s1}(t)$ and $z_{s2}(t)$ included in the observation variable $\boldsymbol{s}(t)$ cannot be measured directly by the sensors. Thus, they should be calculated from the measured data $\ddot{z}_{s1}(t)$ and $\ddot{z}_{s2}(t)$ by numerical integration. The numerical integration can be done by the Newmark-Beta method in Eq. (35). The number of components of $\boldsymbol{Z}(t)$ is 12.

The observability, which is a measure of whether its state variable $\boldsymbol{Z}(t)$ can be estimated from the observed outputs $\boldsymbol{s}(t)$, is expressed by the following condition:

$$\operatorname{rank}\left(\begin{bmatrix} \mathbf{H} \\ \mathbf{HV} \\ \mathbf{HVV} \\ \vdots \\ \mathbf{HV}^{n-1} \end{bmatrix}\right) = n \tag{52}$$

The state-space model of this study satisfies this condition.

The discretized model of Eq. (46) and Eq. (47) can be expressed as

$$\boldsymbol{Z}_k = \overline{\mathbf{V}}\boldsymbol{Z}_{k-1} + \boldsymbol{\omega}_k \tag{53}$$

$$\boldsymbol{s}_k = \mathbf{H}\boldsymbol{Z}_k + \boldsymbol{\epsilon}_k \tag{54}$$

where $\boldsymbol{Z}_k = \boldsymbol{Z}(k\Delta t)$, $\boldsymbol{s}_k = \boldsymbol{s}(k\Delta t)$, $\boldsymbol{\omega}_k = \omega(k\Delta t)$, and $\boldsymbol{\epsilon}_k = \epsilon(k\Delta t)$. $\Delta t$ is the sampling time interval. The matrix $\overline{\mathbf{V}}$ is the matrix exponential of $\mathbf{V}\Delta t$ as following.

$$\begin{aligned} \overline{\mathbf{V}} &= \operatorname{expm}[\mathbf{V}\Delta t] \\ &= \mathbf{U}\operatorname{diag}(\exp(\mathbf{D}))\,\mathbf{U}^{-1} \end{aligned} \tag{55}$$

where $\mathbf{U}$ is the diagonal matrix of which components are the eigenvalues of $[\mathbf{V}\Delta t]$ and $\mathbf{D}$ is the matrix of which columns are the eigenvectors of $[\mathbf{V}\Delta t]$. The operator $\mathrm{diag}(\mathbf{X})$ extract only the diagonal components of $\mathbf{X}$ and returns the diagonal matrix. It can be said that this is an operation to make only the off-diagonal components zero without changing the diagonal components. The eigenvalue decomposition of $[\mathbf{V}\Delta t]$ can be expressed as

$$\mathbf{V}\Delta t = \mathbf{U}\mathbf{D}\mathbf{U}^{-1} \tag{56}$$

The covariance matrix of model error $\boldsymbol{\omega}_k$ and observation error $\boldsymbol{\epsilon}_k$ is $\mathbf{Q}$ and $\mathbf{R}$.

$$\begin{aligned}\mathbf{Q} &= \mathrm{E}[\boldsymbol{\omega}_k\boldsymbol{\omega}_k^{\mathrm{T}}] \\ \mathbf{R} &= \mathrm{E}[\boldsymbol{\epsilon}_k\boldsymbol{\epsilon}_k^{\mathrm{T}}]\end{aligned} \tag{57}$$

where $\mathrm{E}[\ \ ]$ denotes an expected value.

### 3.1. Road Profile Estimation by Karman Filter

In the previous study, Nagayama et al. have succeeded in estimating input profile from vehicle vibration data with noise by Karman Filter. In their study, the vehicle runs only on the ground, and bridges are not considered. In this case, the input profile $\boldsymbol{u}(t)$ equals to the road profile $\boldsymbol{r}(t)$. On the other hand, the vehicle model of this study is affected by the bridge profile $\mathbf{L}(t)^{\mathrm{T}}\boldsymbol{y}(t)$, shown in Eq. (20). However, despite such differences, Nagayama's method is still applicable to the input estimation of this study.

Assume that the estimated state vector is $\hat{\boldsymbol{Z}}_k$. If $\hat{\boldsymbol{Z}}_{k-1}$ is obtained, Eq. (53) gives $\overline{\mathbf{V}}\hat{\boldsymbol{Z}}_{k-1}$ as a candidate for $\hat{\boldsymbol{Z}}_k$, while Eq. (54) also gives $\mathbf{H}^{-1}\boldsymbol{s}_k$ as another candidate. The former is based on Mechanics, while the latter is on data. Both include uncertainties caused by the noises $\boldsymbol{\omega}_k$ and $\boldsymbol{\epsilon}_k$. This means that they can be assumed to be random variables. Let $\hat{\boldsymbol{Z}}_k^{(a)}$ be the former, and if $\hat{\boldsymbol{Z}}_k^{(a)}$ is a standard random variable with mean $\mu_a = \mathrm{E}\left[\hat{\boldsymbol{Z}}_k^{(a)}\right] = \overline{\mathbf{V}}\hat{\boldsymbol{Z}}_{k-1}$, its covariance matrix $\boldsymbol{\Sigma}_a$ can be given by

$$\boldsymbol{\Sigma}_a = \overline{\mathbf{V}}\mathbf{P}_{k-1}\overline{\mathbf{V}}^{\mathrm{T}} + \mathbf{Q} \tag{58}$$

where $\mathbf{P}_{k-1}$ is the posterior covariance matrix of $\hat{\boldsymbol{Z}}_{k-1}$. Similarly, let $\hat{\boldsymbol{Z}}_k^{(b)}$ be the latter, and if $\hat{\boldsymbol{Z}}_k^{(b)}$ is a standard random variable with mean $\mu_b = \mathrm{E}\left[\hat{\boldsymbol{Z}}_k^{(b)}\right] = \mathbf{H}^{-1}\boldsymbol{s}_k$, its covariance matrix $\boldsymbol{\Sigma}_b$ can be given by

$$\boldsymbol{\Sigma}_b = \mathbf{H}^{-1}\mathbf{R}\mathbf{H}^{-\mathrm{T}} \tag{59}$$

Since $\mathbf{H}$ is not regular, $\mathbf{H}^{-1}$ cannot be generally calculated. However, $\boldsymbol{\Sigma}_b$ is always used in the form of $\boldsymbol{\Sigma}_b^{-1}$ $(= \mathbf{H}^{\mathrm{T}}\mathbf{R}\mathbf{H})$, so that is not a problem. The likelihood for $\hat{\boldsymbol{Z}}_k$ can be given as the mean of random value, which is obtained from two predictions $\hat{\boldsymbol{Z}}_k^{(a)}$ and $\hat{\boldsymbol{Z}}_k^{(b)}$.

$$\begin{aligned}\hat{\boldsymbol{Z}}_k &= (\boldsymbol{\Sigma}_a^{-1} + \boldsymbol{\Sigma}_b^{-1})^{-1}(\boldsymbol{\Sigma}_a^{-1}\mu_a + \boldsymbol{\Sigma}_b^{-1}\mu_b) \\ &= [\mathbf{I} - \mathbf{G}\mathbf{H}]\{\overline{\mathbf{V}}\hat{\boldsymbol{Z}}_{k-1}\} + [\mathbf{G}\mathbf{H}]\{\mathbf{H}^{-1}\boldsymbol{s}_k\} \\ &= [\mathbf{I} - \mathbf{G}\mathbf{H}]\{\overline{\mathbf{V}}\hat{\boldsymbol{Z}}_{k-1}\} + \mathbf{G}\boldsymbol{s}_k\end{aligned} \tag{60}$$

where

$$\mathbf{G} = \boldsymbol{\Sigma}_a\mathbf{H}^{\mathrm{T}}(\mathbf{H}\boldsymbol{\Sigma}_a\mathbf{H}^{\mathrm{T}} + \mathbf{R})^{-1} \tag{61}$$

$\mathbf{G}$ is called Karman Gain. The posterior covariance matrix of $\hat{\boldsymbol{Z}}_k$ is

$$\mathbf{P}_k = (\boldsymbol{\Sigma}_a^{-1} + \boldsymbol{\Sigma}_b^{-1})^{-1} \tag{62}$$

The initial value of the posterior covariance matrix $\mathbf{P}_0$ is given by the correct value. **Fig. 10** is a conceptual diagram of Eq. (60).

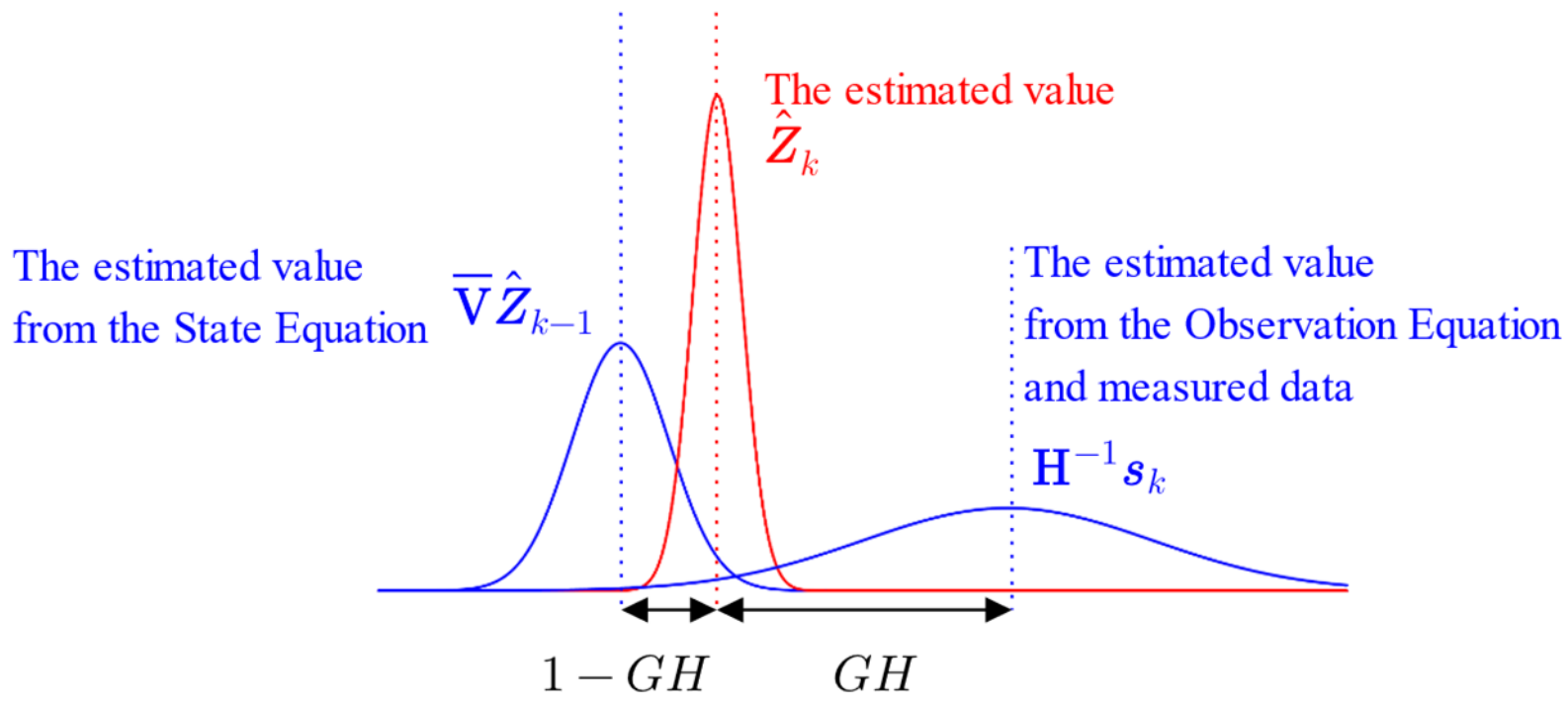

**Fig. 10** The conceptual diagram of estimating for $\hat{\boldsymbol{Z}}_k$ from $\overline{\mathbf{V}}\hat{\boldsymbol{Z}}_{k-1}$ and $\boldsymbol{s}_k$

### 3.2. System Parameter Estimation by Particle Swarm Optimization

The vehicle's and bridge's system parameters are unknown. However, in order to make Karman Filter applicable to the measured data $\boldsymbol{s}_k$ for estimating the input profile $\boldsymbol{u}_k$ $(=\boldsymbol{u}(k\Delta t))$, the system parameters $\overline{\mathbf{V}}$ and $\mathbf{H}$ should be prepared, first. Once all system parameters are assumed, $\boldsymbol{r}_k = \boldsymbol{r}(k\Delta t)$ can be estimated. The system parameter vector $\boldsymbol{X}$ is defined as

$$\boldsymbol{X}_s^{(i)} = \left[m_{s1}^{(i)}\ m_{s2}^{(i)}\ c_{s1}^{(i)}\ k_{s1}^{(i)}\ c_{s2}^{(i)}\ k_{s2}^{(i)}\ m_{u1}^{(i)}\ m_{u2}^{(i)}\ k_{u1}^{(i)}\ k_{u2}^{(i)}\ EI_1^{(i)} \ldots EI_{15}^{(i)}\ \rho A^{(i)}\ \alpha_c\ \beta_c\right]^{\mathrm{T}} \tag{63}$$

where $i$ and $s$ denote the sample number and the step number, respectively. Because $M$ is known, $m_s^{(i)}$ can be calculated from $m_{u1}^{(i)}$ and $m_{u2}^{(i)}$. Similarly, because $D$ is known, once $d_1$ is assumed, $d_2$ can be calculated. The sprung mass $m_{s1}^{(i)}$ and $m_{s2}^{(i)}$ are obtained from $m_s^{(i)}$, $d_1$ and $d_2$. After $\boldsymbol{X}_s^{(i)}$ is assumed, the state vector $\boldsymbol{Z}_k^{(i)}$ and the bridge vibration $\boldsymbol{y}^{(i)}$ can be estimated by Eq. (60) and Eq. (12), respectively. From the input profile $\boldsymbol{u}_k^{(i)}$, which is included in $\boldsymbol{Z}_k^{(i)}$, and the bridge profile $\mathbf{L}^{\mathrm{T}}\boldsymbol{y}^{(i)}$, the road profile $\boldsymbol{r}_k^{(i)}$ $(=\boldsymbol{r}(k\Delta t))$ can be calculated by Eq. (20). By this process, two different road unevenness function $R_1(x)$ and $R_2(x)$ are estimated at the front and rear axles, respectively.

$$\boldsymbol{r}_k^{(i)} = \begin{Bmatrix} r_1^{(i)}(k\Delta t) \\ r_2^{(i)}(k\Delta t) \end{Bmatrix} = \begin{Bmatrix} R_1^{(i)}(x_1(k\Delta t)) \\ R_2^{(i)}(x_2(k\Delta t)) \end{Bmatrix} \tag{64}$$

where $x_1$ and $x_2$ are the positions of the front and rear axles, respectively.

Dividing the position $x$ by $\Delta x$, the objective function can be defined as

$$J\left(\boldsymbol{X}_s^{(i)}\right) = \sum_n \left(R_1(n\Delta x) - R_2(n\Delta x)\right)^2 \tag{65}$$

$R_1$ and $R_2$ are estimates of the same $R$. Thus, if all randomly assumed parameters were the correct values, $J$ could be expected to be zero. Thus, $\boldsymbol{X}_s^{(i)}$ minimizing $J$ should be taken as the estimated system parameters. $\boldsymbol{X}_s^{(i)}$ is repeatedly updated to minimize $J$. If $J\left(\boldsymbol{X}_{s0}^{(i)}\right)$ is the smallest value of the sequence $\left[J\left(\boldsymbol{X}_1^{(i)}\right), J\left(\boldsymbol{X}_2^{(i)}\right), \cdots, J\left(\boldsymbol{X}_{s0}^{(i)}\right), \cdots, J\left(\boldsymbol{X}_s^{(i)}\right)\right]$, $\boldsymbol{X}_{s0}^{(i)}$ is called $i$'s local best:

$$\boldsymbol{L}_s^{(i)} = \boldsymbol{X}_{s_0}^{(i)} \tag{66}$$

Similarly, if $J\left(\boldsymbol{L}_s^{(i0)}\right)$ is the smallest value of the sequence $\left[J\left(\boldsymbol{L}_s^{(1)}\right), J\left(\boldsymbol{L}_s^{(2)}\right), \cdots, J\left(\boldsymbol{L}_s^{(i0)}\right), \cdots, J\left(\boldsymbol{L}_s^{(N)}\right)\right]$, $\boldsymbol{L}_s^{(i0)}$ is called the global best:

$$\boldsymbol{G}_s = \boldsymbol{L}_s^{(i0)} \tag{66}$$

$\boldsymbol{X}_s^{(i)}$ is updated by the following:

$$\boldsymbol{X}_{s+1}^{(i)} = \boldsymbol{X}_s^{(i)} + \Delta\boldsymbol{X}_s^{(i)} \tag{67}$$

where

$$\Delta\boldsymbol{X}_s^{(i)} = \alpha_1 q_{s1}^{(i)} \Delta\boldsymbol{X}_{s-1}^{(i)} + \alpha_2 q_{s2}^{(i)} \left(\boldsymbol{L}_s^{(i)} - \boldsymbol{X}_s^{(i)}\right) + \alpha_3 q_{s3}^{(i)} \left(\boldsymbol{G}_s - \boldsymbol{X}_s^{(i)}\right) \tag{68}$$

The coefficients $\alpha_1$, $\alpha_2$ and $\alpha_3$ are used to adjust the speed of convergence, and $q_{sj}^{(i)}$ is a uniform random value ranging from 0 to 1.

## 3.3. Hyper-parameter setting

The proposed method is a hybrid method: one is PSO method to assume system parameters, and another is Karman filter to estimate the input profile of the vehicle. The accuracy of Karman filter depends on the covariance matrix $\mathbf{R}$ and $\mathbf{Q}$. Since the components of these matrices are unknown, it is necessary to assume them as hyperparameters. **Table 4** and **Table 5** show the covariance of $\mathbf{R}$ and $\mathbf{Q}$, respectively.

**Table 4** The assumed covariance of the system noise $\boldsymbol{\omega}_k$

**(a)** without noise

| $i$ | 1 | 2 | 3 | 4 | 5 | 6 | 7 | 8 | 9 | 10 | 11 | 12 |
|---|---|---|---|---|---|---|---|---|---|---|---|---|
| $Q_{ii}$ | 0 | 0 | 0 | 0 | 0 | 0 | 0 | 0 | 181 | 178 | 133000 | 133000 |

$\times 10^{-7}$

**(b)** with 15% noise

| $i$ | 1 | 2 | 3 | 4 | 5 | 6 | 7 | 8 | 9 | 10 | 11 | 12 |
|---|---|---|---|---|---|---|---|---|---|---|---|---|
| $Q_{ii}$ | 6.00 | 6.00 | 6.00 | 6.00 | 4.50 | 4.50 | 4.50 | 4.50 | 181 | 178 | 133000 | 133000 |

$\times 10^{-7}$

**(c)** with 35% noise

| $i$ | 1 | 2 | 3 | 4 | 5 | 6 | 7 | 8 | 9 | 10 | 11 | 12 |
|---|---|---|---|---|---|---|---|---|---|---|---|---|
| $Q_{ii}$ | 60.0 | 60.0 | 60.0 | 60.0 | 45.0 | 45.0 | 45.0 | 45.0 | 181 | 178 | 133000 | 133000 |

$\times 10^{-7}$

**Table 5** The assumed covariance of the observation noise $\epsilon_k$

**(a)** without noise

| $i$ | 1 | 2 | 3 | 4 |
|---|---|---|---|---|
| $R_{ii}$ | 1.00 | 1.00 | 1.00 | 1.00 |

$\times 10^{-9}$

**(b)** with 15% noise

| $i$ | 1 | 2 | 3 | 4 |
|---|---|---|---|---|
| $R_{ii}$ | 6.00 | 6.00 | 7.50 | 7.50 |

$\times 10^{-4}$

**(c)** with 35% noise

| $i$ | 1 | 2 | 3 | 4 |
|---|---|---|---|---|
| $R_{ii}$ | 6.00 | 6.00 | 7.50 | 7.50 |

$\times 10^{-3}$

The coefficients of PSO method, $\alpha_1$, $\alpha_2$ and $\alpha_3$ are also hyperparameters of this method. The used values are shown in **Table 6**. Since the PSO method randomly assumes and updates several combinations of target parameters, the sample number is an important parameter. After it updates, the PSO method returns the best combination, despite the number of samples. However, if large samples are prepared, they must consume too many computing resources. Thus, in this study, results are presented in probability by executing vehicle runs repeatedly.

**Table 6** The coefficient of PSO method

| Samples | Runs | $\alpha_1$ | $\alpha_2$ | $\alpha_3$ |
|---|---|---|---|---|
| 60 | 100 | 0.60 | 0.30 | 0.10 |

$\times 10^{-9}$

The initial value for each parameter, except $d_1$ and damaged $EI$, is assumed randomly in the range from 0.8 to 1.2 times the correct value. For the flexural rigidity $EI$, the assumed values are generated from 0.8 to 1.2 times the correct intact value. The distance $d_1$ is randomly assumed from 0.1 to 0.9 times the distance $D$. This means that the inspector knows the approximate values of the system parameters in advance.

## 3.4. Results and Discussion

### (a) The cases without noises

The proposed method estimates 1) the vehicle's system parameters, 2) the bridge's system parameters, and 3) the road unevenness at the same time. **Fig. 11** shows the estimated values of the vehicle parameters for the case of "intact bridge" and "without noise". The parameter estimation is repeated 100 times. The height of bars is normalized so that its total area equals one, which means that the vertical axis indicates the probability. The horizontal axis is also normalized to denote the ratio of estimated value to the correct value, which means that 1 indicates its correct value. The blue and red bars indicate the prior and posterior distributions, respectively. The prior distribution indicates that the initial distribution before applying the PSO method. The red bars indicate the distribution of global bests of all run cases. By comparison of blue and red bars, it can be said that the PSO method can make the accuracy of estimating parameters better than the random assumption.

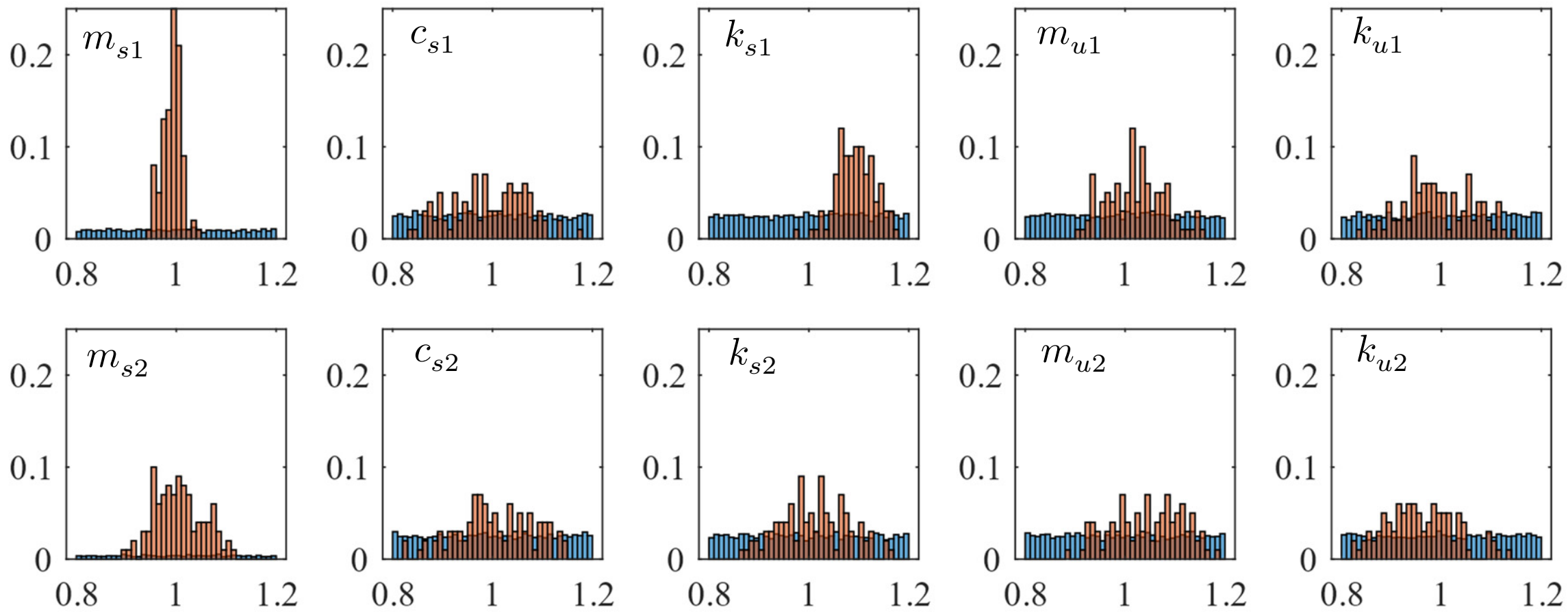

**Fig. 11** The estimated normalized system parameters of the vehicle (blue: prior dist., red: posterior dist.)

According to **Fig. 11**, the estimation accuracies for sprung-mass $m_{s1}$ and $m_{s2}$ are quite high. Comparing the estimation accuracies of $m_{s1}$ and $m_{s2}$, the former is better than the latter. The distance from the gravity point to the observation point may affect the result. The estimation accuracies for the parameters of suspensions ($c_{s1}$, $c_{s2}$, $k_{s1}$, and $k_{s2}$) and tires ($m_{u1}$, $m_{u2}$, $k_{u1}$, and $k_{u2}$) are not as high as those of vehicle body ($m_{s1}$ and $m_{s2}$). However, they should be appropriately updated by the PSO method from the initial prior distribution.

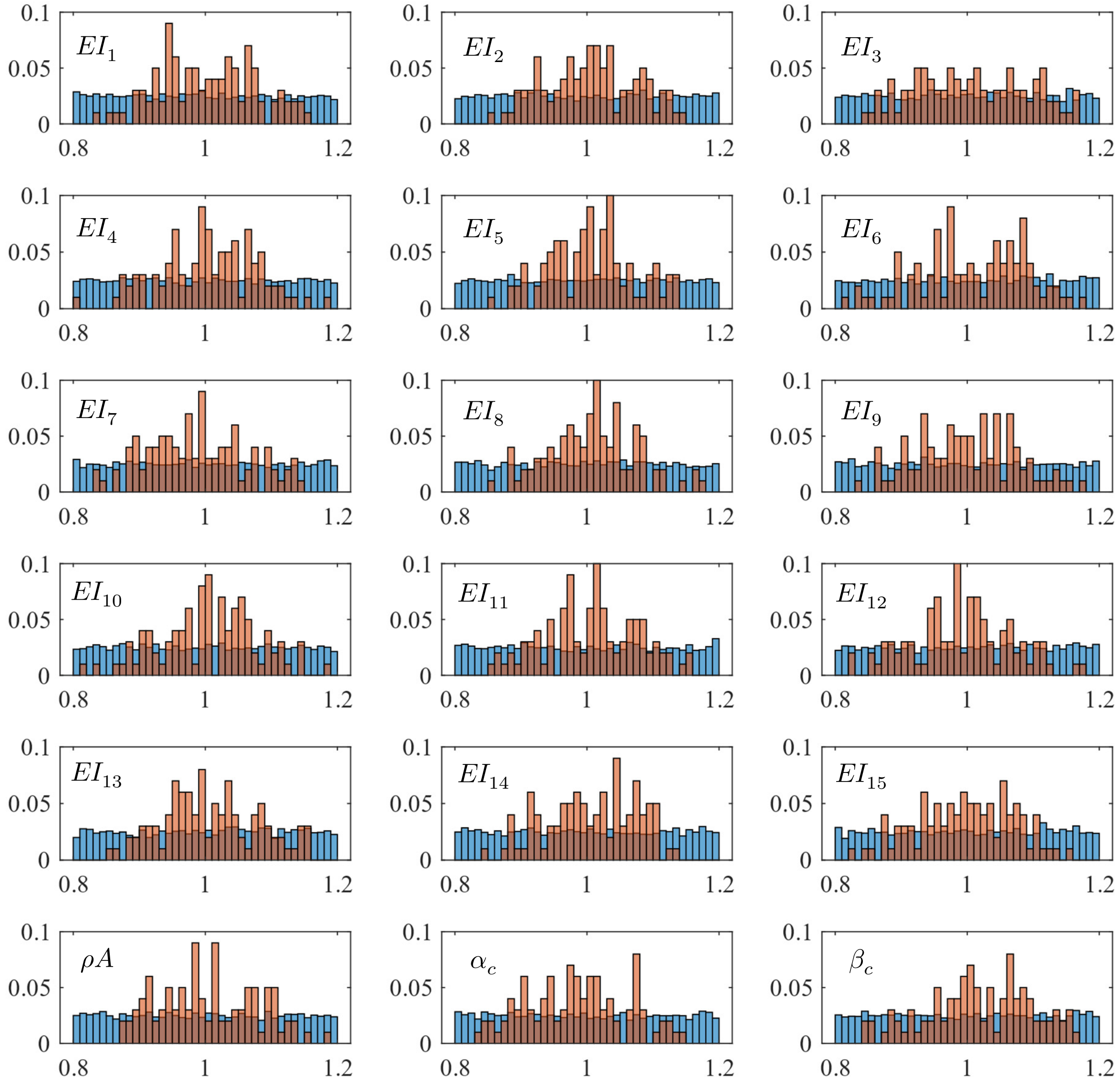

**Fig. 12** The estimated normalized system parameters of the bridge
(blue: prior dist., red: posterior dist.)

**Fig. 12** shows the estimated values of the bridge for the same case. The subscript of $EI$ denotes an element number. According to this figure, the estimation accuracy of $EI$ at mid-span ($EI_8$ is the center of the bridge) is relatively higher than that near the edges ($EI_1$ and $EI_{15}$). The cause of the decrease in estimation accuracy is that the pin supports restrain the bridge deflections at the edges. Because the vibration is slight, the accuracy at each edge is relatively lower. The mass $\rho A$ of the bridge can also be estimated. On the other hand, the damping coefficients $\alpha_c$ and $\beta_c$ are not accurately estimated because their peaks are different from the correct values.

These results indicate that some parameters are susceptible and others are not. On the other hand, it is also confirmed that this method is an effective method for estimating the mechanical parameters of vehicles and bridges.

**Fig. 13** shows the estimated road unevenness. The blue and red curves denote the estimated ones at the front and rear axles, while the black curve is the correct. In this figure, both estimated road unevenness looks similar to the correct one. Thus, the Karman filter is efficient in estimating vehicle inputs.

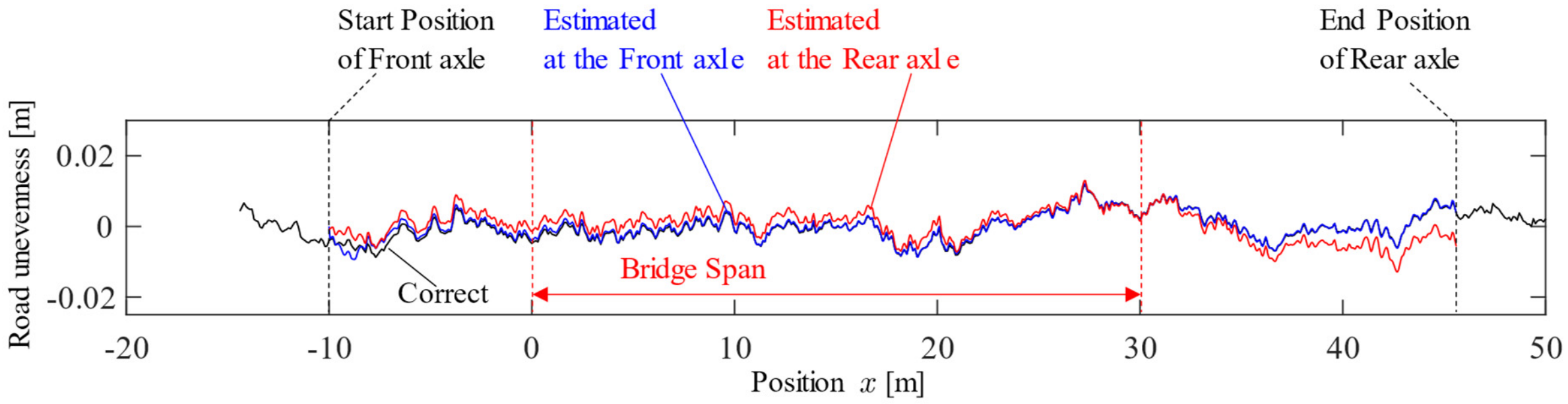

**Fig. 13** The estimated road unevenness

However, though there is no noise, the error can also be observed. There is a limitation in estimating the system parameters by this method. According to **Fig. 11**, **Fig. 12** and **Fig. 13**, the estimation accuracy at the front axle is better than that at the rear axle. There is a possibility that the distance of the observation point from the gravity point affects the accuracy. Because the tendency of error included in the estimated road unevenness seems to be a low-frequency component, the accuracy decline due to the numerical integration should be concerned.

## (b) The cases with 15% noises

**Fig. 14** shows the estimated values of the vehicle parameters for the case of "intact bridge" and "with 15% noise". Due to the noise, the estimation accuracy becomes lower than the "without noise" cases.

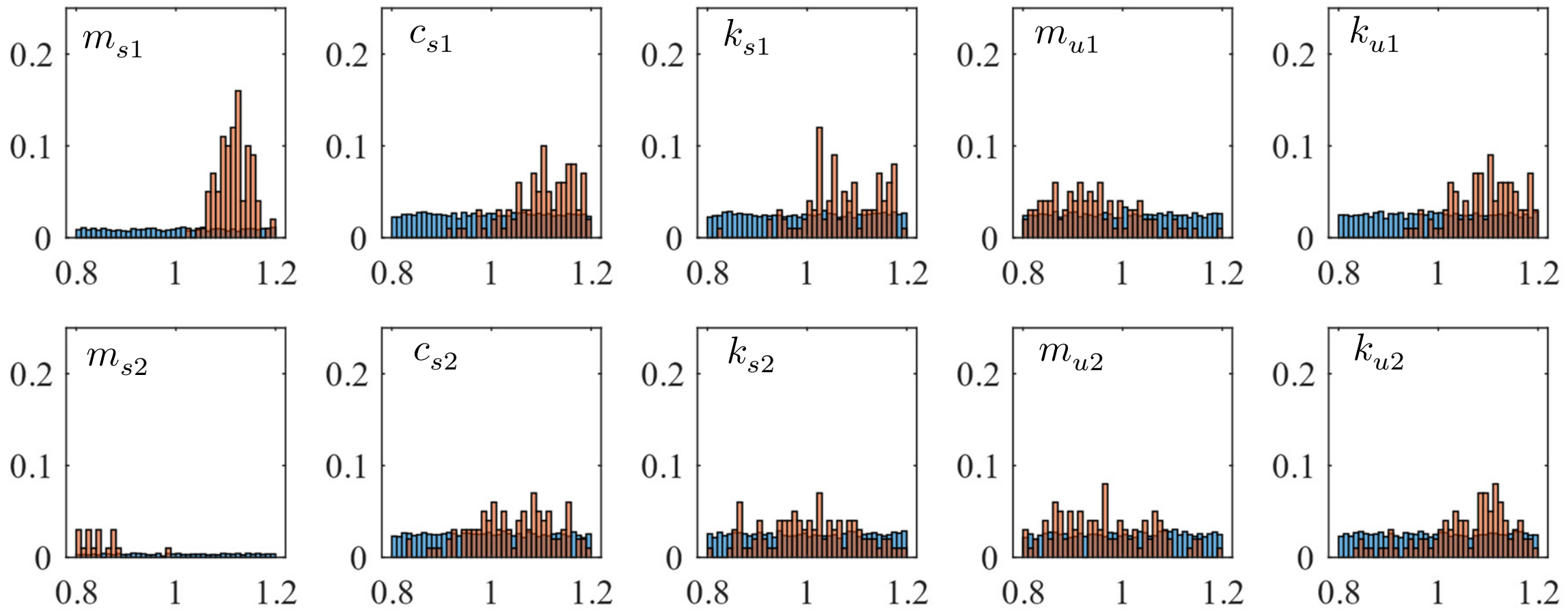

**Fig. 14** The estimated normalized system parameters of the vehicle

(blue: prior dist., red: posterior dist.)

**Fig. 15** also shows the estimated values of the bridge parameters for the same case. Similarly, the accuracy becomes lower than the "without noise" case.

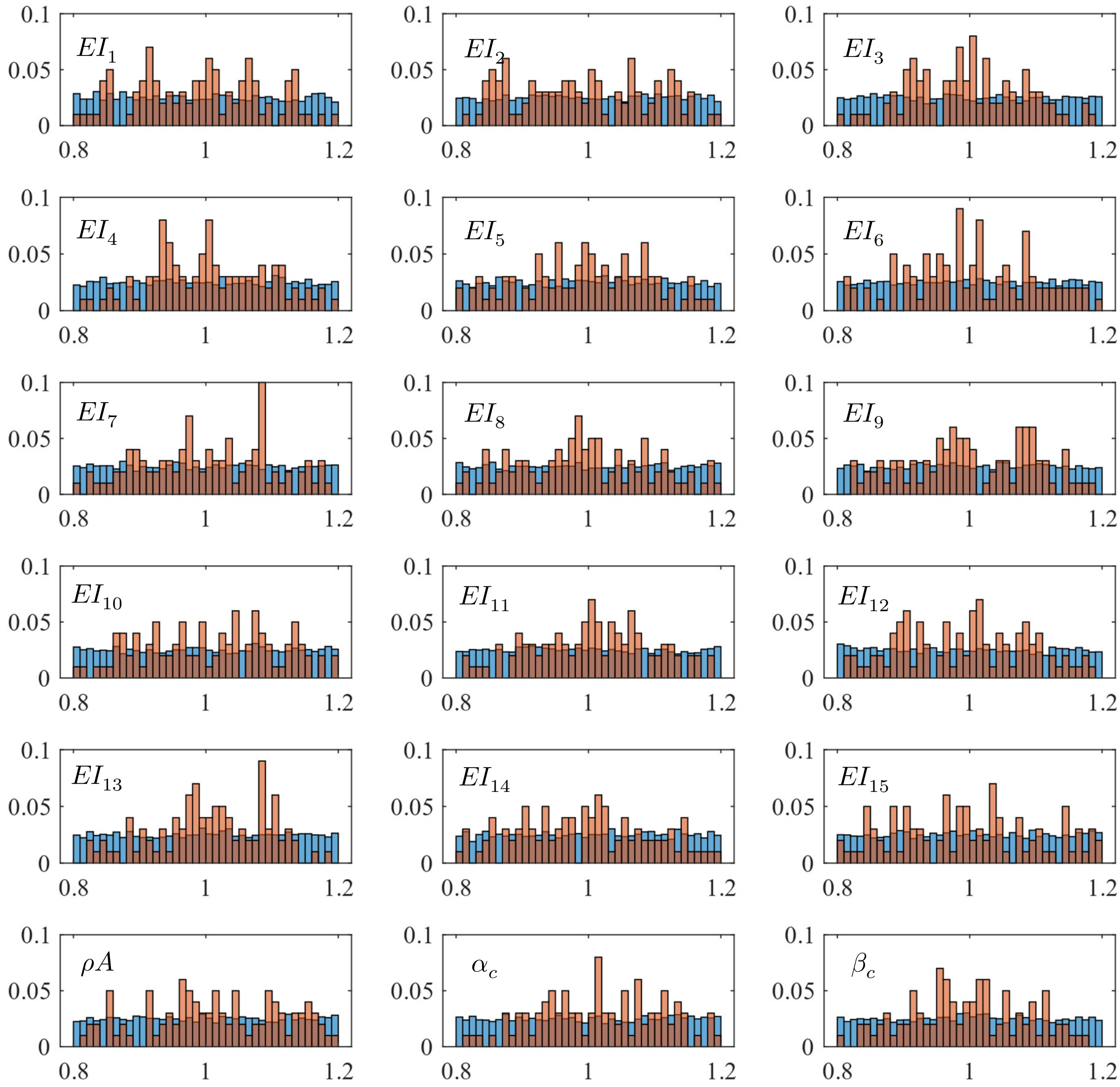

**Fig. 15** The estimated normalized system parameters of the bridge

(blue: prior dist., red: posterior dist.)

**Fig. 16** shows the estimated road unevenness. The estimated waves of road unevenness are not so different from the correct values. However, they include high-frequency errors due to noise and the trend error of the estimated one at the rear axle increases.

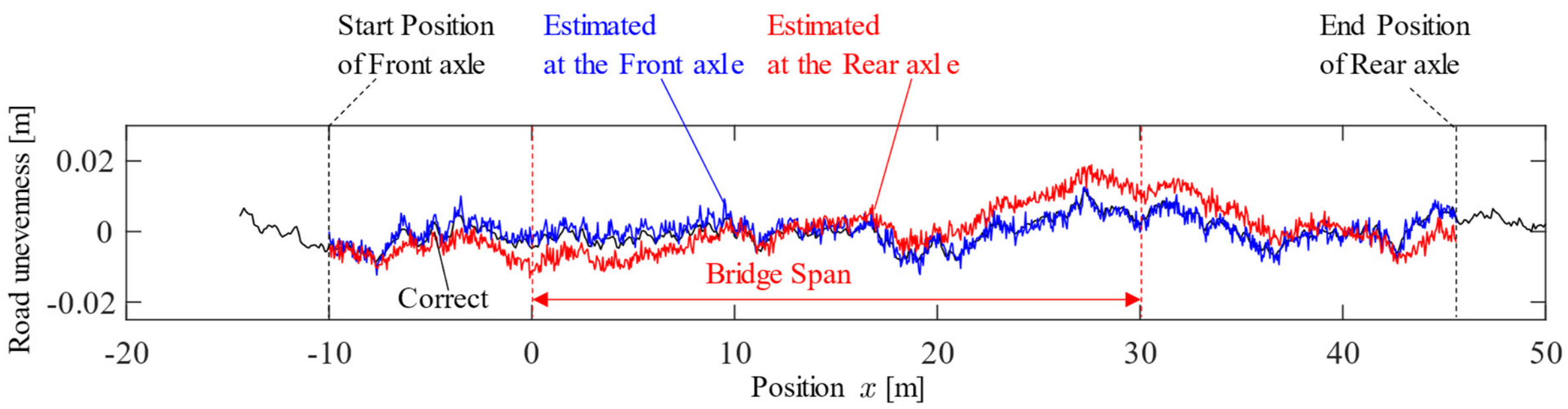

**Fig. 16** The estimated road unevenness

### (c) The cases with 35% noises

**Fig. 17** and **Fig. 18** show the estimated values of the vehicle and bridge parameters for the case of “intact bridge” and “with 35% noise”. This case shows the same tendency as the “with 15% noise” case. As observation noises increase, the estimation accuracy becomes dependent on prior information.

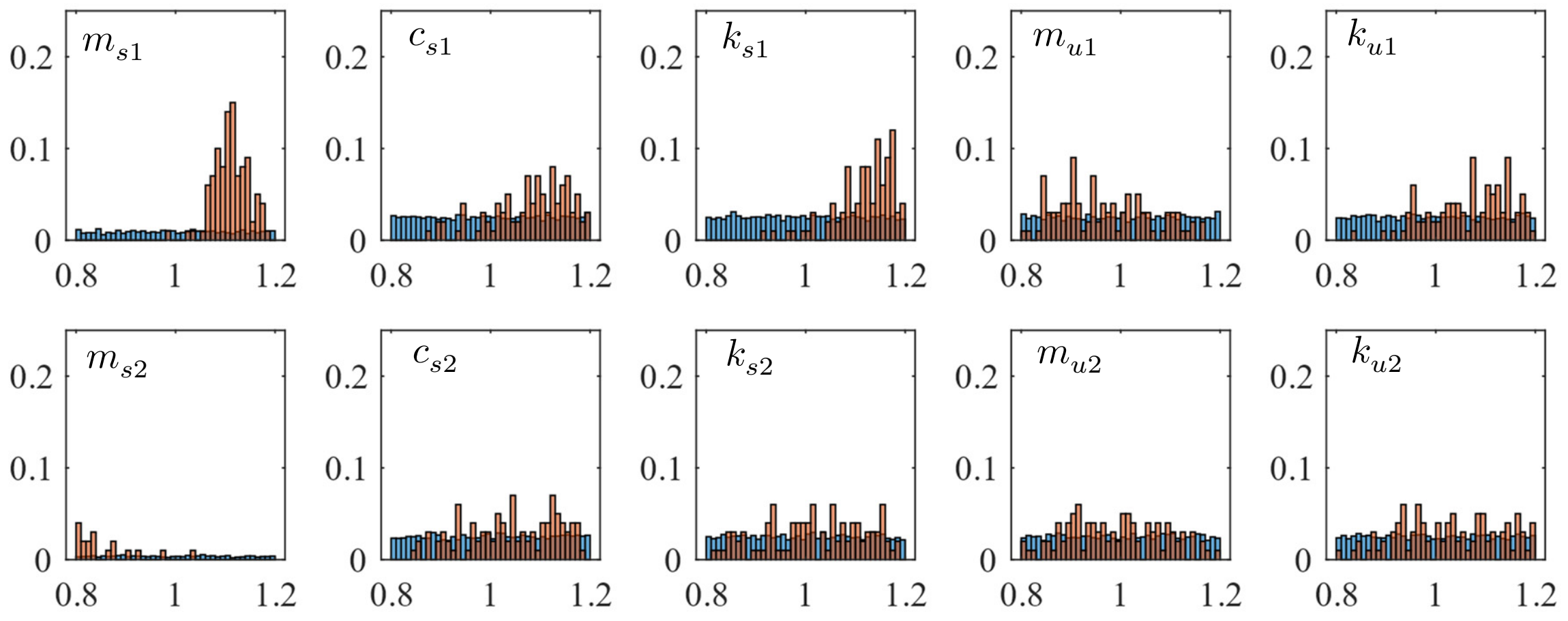

**Fig. 17** The estimated normalized system parameters of the vehicle

(blue: prior dist., red: posterior dist.)

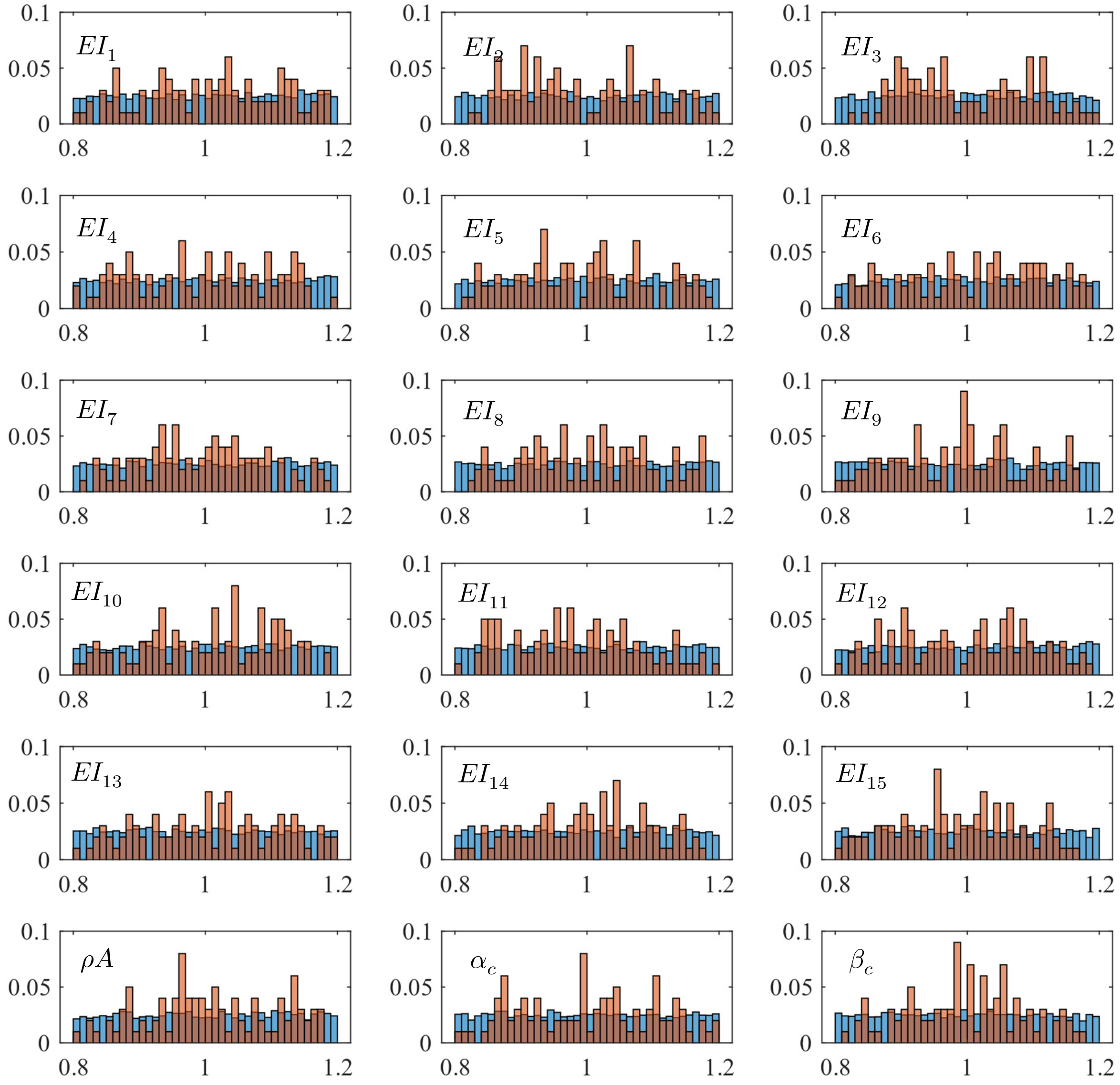

**Fig. 18** The estimated normalized system parameters of the bridge
(blue: prior dist., red: posterior dist.)

**Fig. 19** shows the estimated road unevenness of the "with 35% noise" case. The estimation accuracy of the rear axle becomes worse than the other cases. The cause of the low estimation accuracy for the system parameters should be the low accuracy in estimating road profiles by the Karman filter.

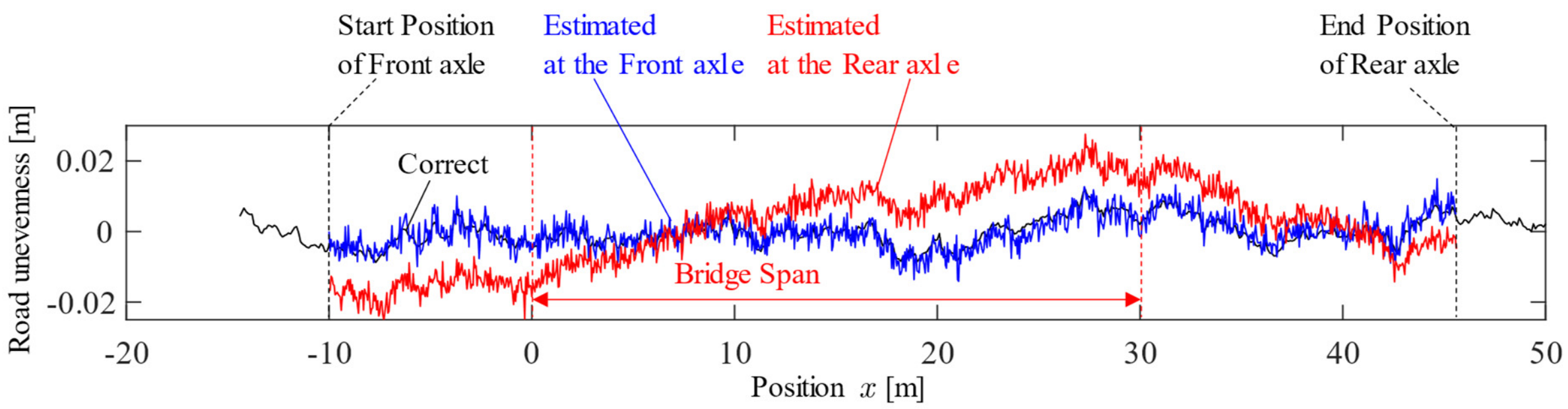

**Fig. 19** The estimated road unevenness

## (d) The cases of damaged bridge

**Fig. 20** and **Fig. 21** show the estimated values of the vehicle and bridge parameters for the case of "damaged bridge" and "without noise". **Fig. 22** also shows the estimated road unevenness. According to these figures, the "damaged bridge" case shows the same result as the "intact bridge" case. However, The correct value of $EI_8$ at the center of the span is 0.5, indicating that the proposed method fails to detect the damage. Although the proposed method can identify all the system parameters, it is not accurate enough to be used for vehicle and bridge inspection, which indicates that it should be improved.

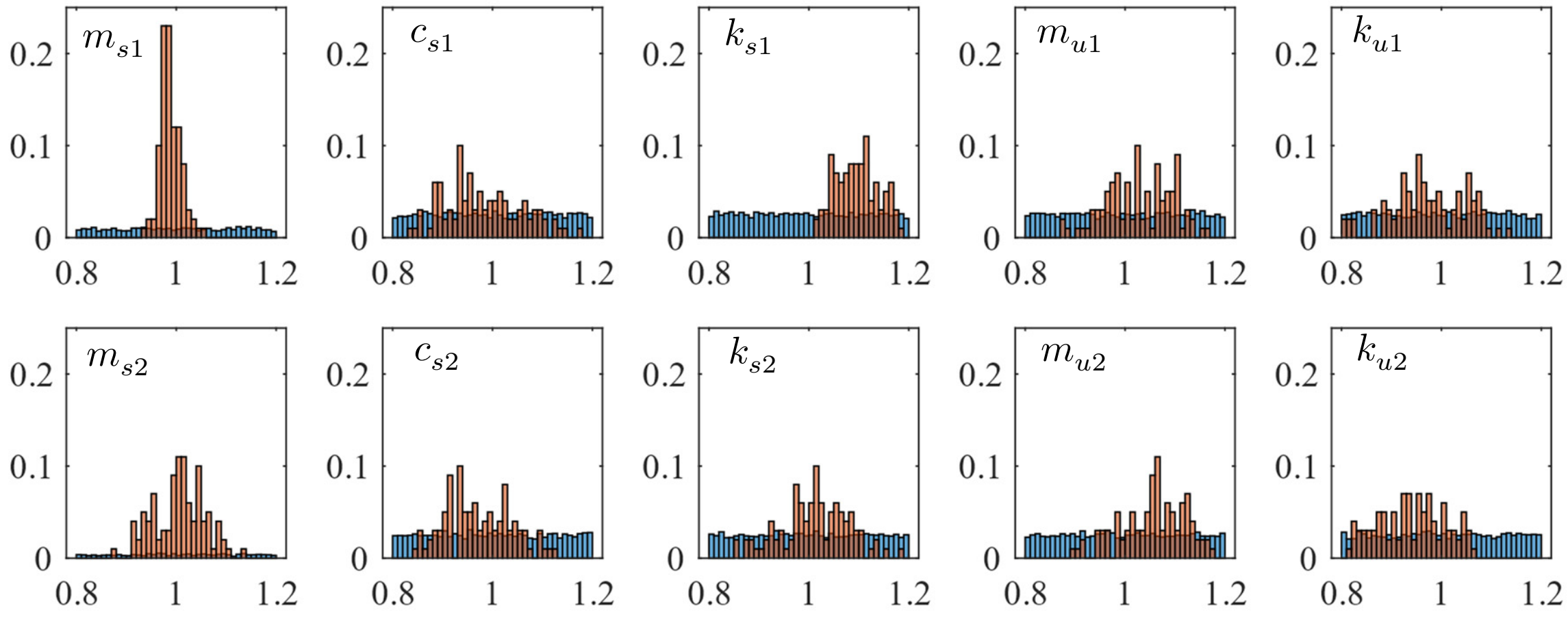

**Fig. 20** The estimated normalized system parameters of the vehicle

(blue: prior dist., red: posterior dist.)

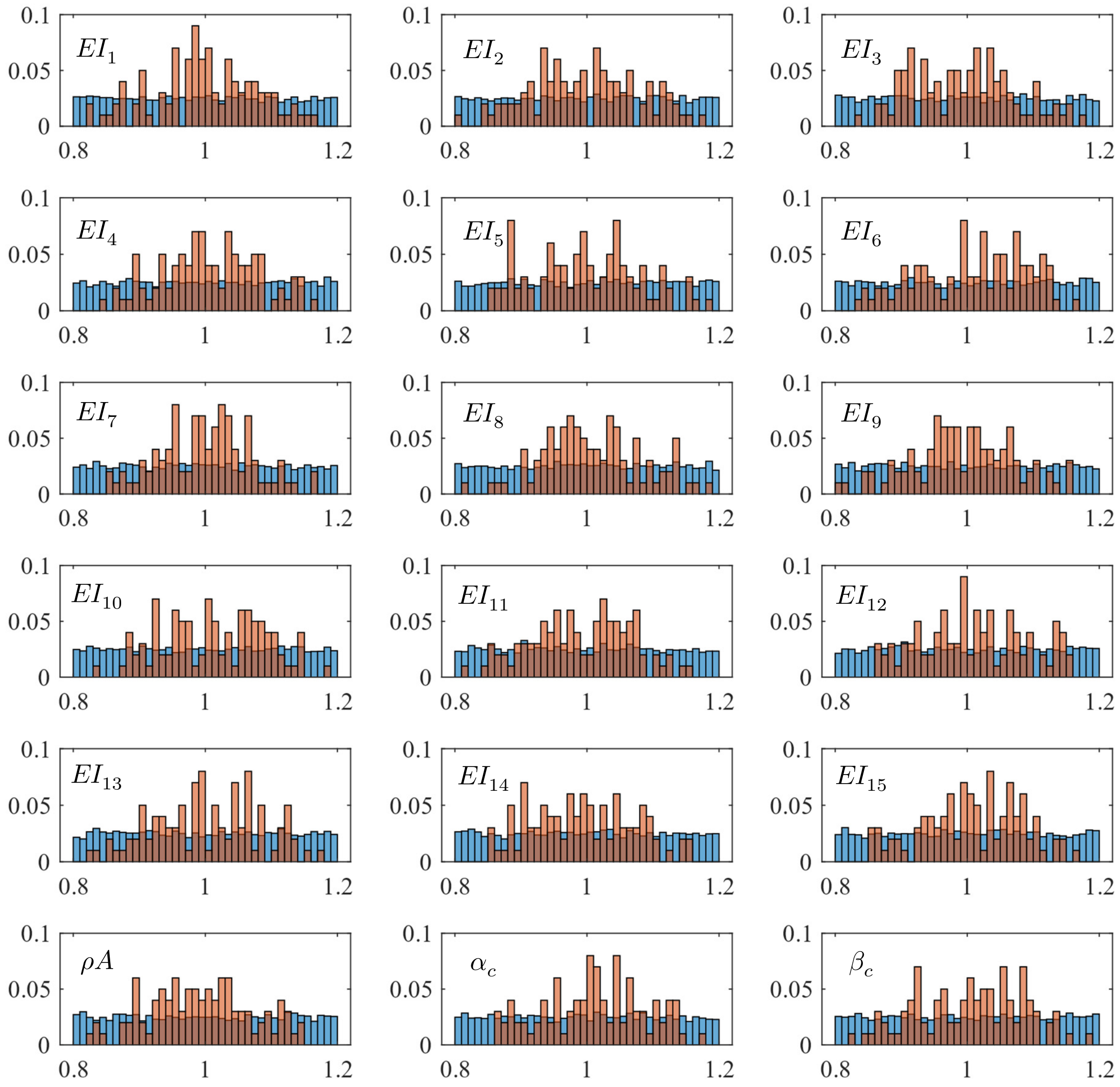

**Fig. 21** The estimated normalized system parameters of the bridge
(blue: prior dist., red: posterior dist.)

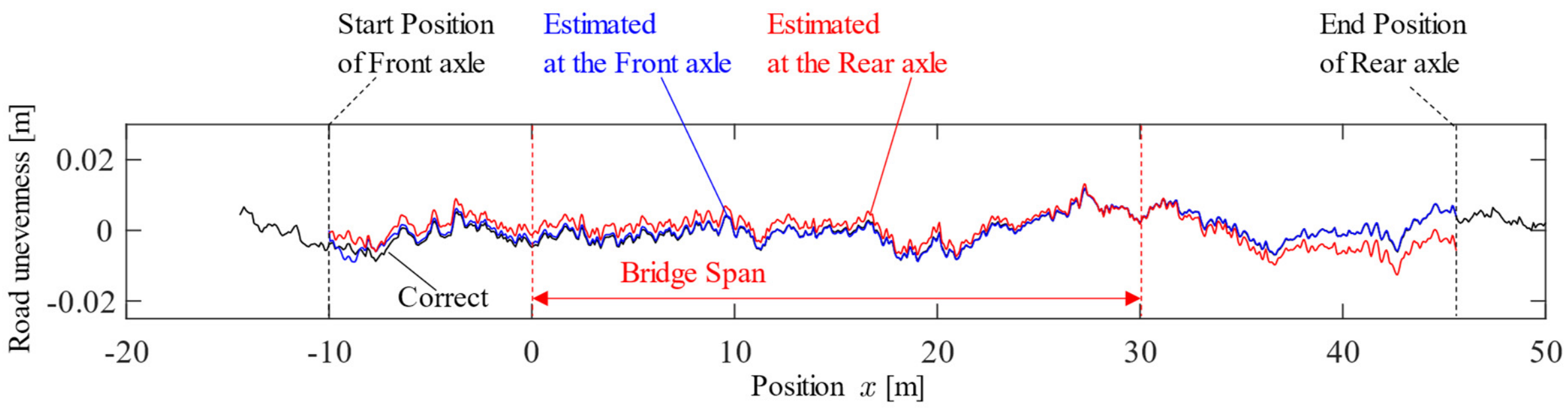

**Fig. 22** The estimated road unevenness

## 4. Conclusion

This study proposes a method to identify the vehicle's and bridge's system parameters and estimate road unevenness only from the vehicle vibration data. The proposed method consists of the PSO method and the Karman filter. The systems parameters are assumed randomly first and updated by the PSO method based on the road profiles estimated by the Karman filter.

The numerical experiments show that this method can estimate the system parameters and road unevenness with better accuracy than random assumption. However, the observation noises affect the accuracy of road profile estimation. The low accuracy of input estimation becomes the cause of the low accuracy of parameter estimation. This method cannot be applied for vehicle and bridge inspection due to the low estimation accuracy. This method should be improved.

The PSO method is not appropriate because it is a Monte Carlo-like method and consumes computational resources. In the "without noise" case, the randomly assumed prior parameters tended to move in the direction to the correct value, so it can be expected that the objective function is less multimodal. If the prior parameters can be set in the range of its low multimodality, a fast algorithm can be designed to replace the PSO method. After implementing the fast algorithm, the low estimation accuracy due to the noise should be addressed.

In this study, a normal probability distribution is assumed for the noise. However, the whiteness of the noise cannot be assumed in an actual vehicle because of the engine vibration. Since noise has a significant effect on the results, the noise characteristics should be experimentally understood.

## Acknowledgement

This study is supported by 1) **MEXT/JSPS KAKENHI Grant Number JP19H02220** and 2) **Shoreki Commemorative Foundation**.

## Appendix

### A. Mass and Stiffness matrices of the bridge FE model

The mass matrix of bridge can be written by element mass matrices, in the form of

$$\mathbf{M}_b = \left[\int_0^L \rho A \boldsymbol{N}\boldsymbol{N}^{\mathrm{T}} \mathrm{d}x\right] = \sum_{j=1}^{n} \left[\int_{x_j}^{x_{j+1}} \rho A \boldsymbol{N}\boldsymbol{N}^{\mathrm{T}} \mathrm{d}x\right] = \sum_{j=1}^{n} \mathbf{M}_b^{(j)} \tag{A.1}$$

where $\mathbf{M}_b^{(j)}$ is the $j$-th element mass matrix. As the components of this matrix that are corresponding to the degrees of freedom of the $j$-th element is defined as submatrix $\mathbf{m}_b^{(j)}$. Similarly, the stiffness matrix can be also written in

$$\mathbf{K}_b = \left[\int_0^L EI \frac{\partial^2 \boldsymbol{N}}{\partial x^2}\frac{\partial^2 \boldsymbol{N}^{\mathrm{T}}}{\partial x^2} \mathrm{d}x\right] = \sum_{j=1}^{n} \left[\int_{x_j}^{x_{j+1}} EI \frac{\partial^2 \boldsymbol{N}}{\partial x^2}\frac{\partial^2 \boldsymbol{N}^{\mathrm{T}}}{\partial x^2} \mathrm{d}x\right] = \sum_{j=1}^{n} \mathbf{K}_b^{(j)} \tag{A.2}$$

where $\mathbf{K}_b^{(j)}$ is the $j$-th element mass matrix. The submatrix corresponding to the degrees of freedom of the $j$-th element of $\mathbf{K}_b^{(j)}$ is $\mathbf{k}_b^{(j)}$. The Eqs (A.1) and (A2) represent the principle of superposition.

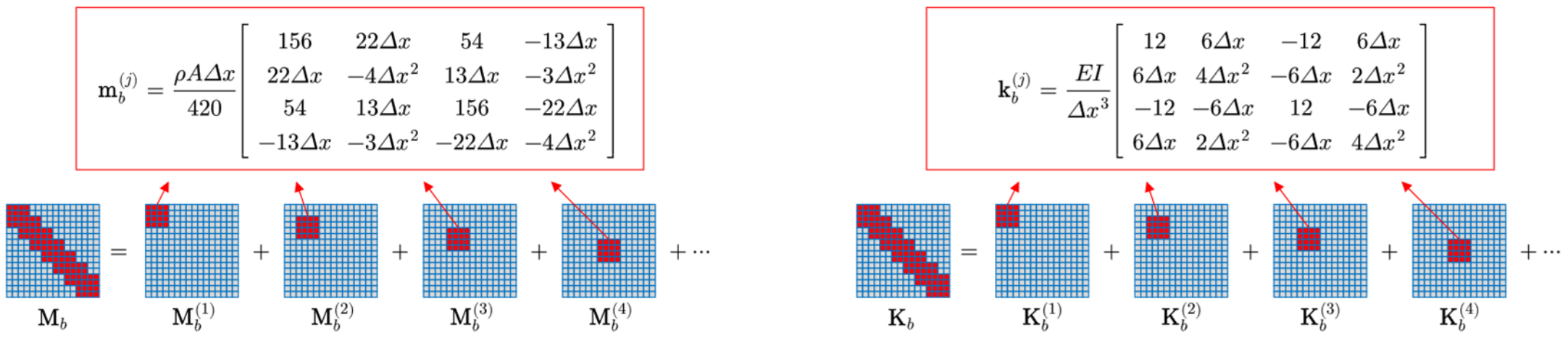

(a) Mass matrix (b) Stiffness matrix

**Fig. A-1** The principle of superposition